\documentclass[letterpaper,twoside,journal]{IEEEtran}
\usepackage{ifpdf}
\usepackage{color}

\usepackage{cite}
\ifCLASSINFOpdf
 \usepackage[pdftex]{graphicx}
 \graphicspath{{../pdf/}{../jpeg/}}
 \DeclareGraphicsExtensions{.pdf,.jpeg,.png}
\else
 \usepackage[dvips]{graphicx}
 \usepackage[dvips]{epsfig}
 \graphicspath{{../eps/}}
 \DeclareGraphicsExtensions{.eps}
\fi
\usepackage{amsmath}
\usepackage{array}
\usepackage{booktabs}
\usepackage{multirow}
\usepackage{multicol}
\usepackage{amssymb}

\usepackage{lipsum} % To generate some text for the article via \lipsum

\usepackage{tikz,xcolor,hyperref}

\definecolor{lime}{HTML}{A6CE39}
\DeclareRobustCommand{\orcidicon}{%
	\begin{tikzpicture}
	\draw[lime, fill=lime] (0,0) 
	circle [radius=0.16] 
	node[white] {{\fontfamily{qag}\selectfont \tiny ID}};
	\draw[white, fill=white] (-0.0625,0.095) 
	circle [radius=0.007];
	\end{tikzpicture}
	\hspace{-2mm}
}

\foreach \x in {A, ..., Z}{%
	\expandafter\xdef\csname orcid\x\endcsname{\noexpand\href{https://orcid.org/\csname orcidauthor\x\endcsname}{\noexpand\orcidicon}}
}

\begin{document}
\title{Most Efficient Sensor Network Protocol for a Permanent Natural Disaster Monitoring System}

\author{Changmin Lee\orcidA{},~\IEEEmembership{Student Member,~IEEE, and}
 Seong-Lyun Kim\orcidB{},~\IEEEmembership{Member,~IEEE}% <-this % stops a space
\thanks{Manuscript received July 20, 2020; revised December 20, 2020; revised January 30, 2021; accepted February 28, 2021; Date of publication -,-; date of current version -,-; This research was supported by a grant to Bio-Mimetic Robot Research Center Funded by Defense Acquisition Program Administration, and by Agency for Defense Development (UD190018ID). A part of this paper was presented at the European Conference on Networks and Communications in 2020 and was published in its Proceedings (DOI:10.1109/EuCNC48522.2020.9200960). \textit{(Corresponding author: Seong-Lyun Kim.)}}
\thanks{Changmin Lee are with the Department
of Electrical and Electronic Engineering, Yonsei University, Seoul,
Korea e-mail: cmlee@ramo.yonsei.ac.kr.}% <-this % stops a space
\thanks{Seong-Lyun Kim are with the Department
of Electrical and Electronic Engineering, Yonsei University, Seoul, Korea e-mail: slkim@yonsei.ac.kr.}% <-this % stops a space
}

%   ^------------^------------^----Do not want these spaces!

\markboth{IEEE Internet of Things,~Vol.~X, No.~X, X~2021}{Lee \MakeLowercase{\textit{et al.}}: Most Efficient Sensor Network Protocol for a Permanent Natural Disaster Monitoring System}

% *** name in the headers of peer review papers.   ***

\IEEEpubid{\begin{minipage}{\textwidth}\ \\ \centering 2327--4662~\copyright~2021 IEEE. Personal use is permitted, but republication/redistribution requires IEEE permission. \\ See http://www.ieee.org/publications\_standards/publications/rights/index.html for more information.\end{minipage}}


\maketitle

\begin{abstract}
To minimize enormous havoc from disasters, permanent environment monitoring is necessarily required. Thus we propose a novel energy management protocol for energy harvesting wireless sensor networks (EH-WSNs), named the \emph{adaptive sensor node management protocol (ASMP)}. The proposed protocol makes system components to systematically control their performance to conserve the energy. Through this protocol, sensor nodes autonomously activate an additional energy conservation algorithm. ASMP embeds three sampling algorithms. For the optimized environment sampling, we proposed the \emph{adaptive sampling algorithm for monitoring (ASA-m)}. ASA-m estimates the expected time period to occur meaningful change. The meaningful change refers to the distance between two target data for the monitoring QoS. Therefore, ASA-m merely gathers the data the system demands. The \emph{continuous adaptive sampling algorithm (CASA)} solves the problem to be continuously decreasing energy despite of ASA-m. When the monitored environment shows a linear trend property, the sensor node in CASA rests a sampling process, and the server generates predicted data at the estimated time slot. For guaranteeing the self-sustainability, ASMP uses the \emph{recoverable adaptive sampling algorithm (RASA)}. RASA makes consumed energy smaller than harvested energy by utilizing the predicted data. RASA recharges the energy of the sensor node. Through this method, ASMP achieves both energy conservation and service quality.
\end{abstract}

\begin{IEEEkeywords}
Energy Harvesting Wireless Sensor Networks, Energy Conservation Protocol, Adaptive Sampling Algorithm, Disaster Monitoring System, Sampling Period Prediction, Self-sustainability
\end{IEEEkeywords}

% \fi
\IEEEpeerreviewmaketitle

\section{Introduction}
%\IEEEPARstart{T}{he} permanent monitoring system requires key features, such as self-sustainability, Quality of Service (QoS), and low complexity. To establish this monitoring system, we proposes a protocol and three energy conservation algorithms, \emph{adaptive sensor node management protocol (ASMP)}, \emph{adaptive sampling algorithm for monitoring (ASA-m)}, \emph{continuous adaptive sampling algorithm (CASA)}, and \emph{recoverable adaptive sampling algorithm (RASA)}. We have summarized the key contents of all proposed works as follows.
\IEEEPARstart{T}{his} study proposes a protocol and three energy conservation algorithms, \emph{adaptive sensor node management protocol (ASMP)}, \emph{adaptive sampling algorithm for monitoring (ASA-m)}, \emph{continuous adaptive sampling algorithm (CASA)}, and \emph{recoverable adaptive sampling algorithm (RASA)}, to ensure three requirements for the permanent monitoring system, such as self-sustainability, Quality of Service (QoS)\footnote{We define QoS as a spatial resolution of data collected on a server-side, following definitions by Zhang \textit{et al.}\cite{Zhang2005} and Al-Shammari \textit{et al.}\cite{Al-Shammari2018}.}, and low complexity. We summarize key contents of all proposed works as follows. 

\subsection{A brief summary of the adaptive sensor node management protocol (ASMP)}
ASMP is a network protocol to embed three proposed algorithms. ASMP switches three proposed algorithms according to the energy condition of a sensor node. To control this dynamic system, ASMP classifies sensor nodes into three classes, A, B, or C. Class A is a state in QoS optimized operation. Class B is a state for forcibly limiting the data gathering ability of sensor nodes for energy conservation. The operations in each class are defined as follows:
\begin{itemize}
  \item Class A: the state activating ASA-m.
  \item Class B: (Class 1) the state activating ASA-m and CASA.
  \item \quad\quad\quad\;\, (Class 2) the state activating ASA-m and RASA.
  \item Class C: the state in sleep state.
\end{itemize}

\IEEEpubidadjcol

\subsubsection{Adaptive Sampling Algorithm for monitoring (\textbf{ASA-m})}
ASA-m estimates the next sampling period optimized to QoS requirements. QoS means the spatial resolution that an application demands. ASA-m picks out a certain value to satisfy the resolution from a monitoring environment. This process runs both on sensor nodes and the server in a real-time manner. ASA-m reduces up to $65.4\%$ of energy consumption and $50\%$ of sampling counts compared to the fixed sampling. 

\subsubsection{Continuous Adaptive Sampling Algorithm (\textbf{CASA})} 
CASA makes a sensor node skip one sampling operation when the monitoring signal linearly changes. The server generates a predicted data to complement the QoS reduction. To support this operation, the sensor node shares its class information with the server. Through this approach, the sensor node reduces up to $83.3\%$ of the consumption energy and $60\%$ of the sampling counts compared to the fixed sampling. 

\subsubsection{Recoverable Adaptive Sampling Algorithm (\textbf{RASA})}
RASA recovers the energy state of sensor nodes when they are in danger of energy depletion. It is a key algorithm to guarantee self-sustainability. RASA makes energy consumption lower than the harvested by extending a sampling period. RASA sampling period extends to ASA-m period times RASA factor ($\textbf{N}$). To satisfy QoS requirements, the server generates $(\mathbf{N}-1)$ numbers of predicted data at ASA-m sampling periods.

\subsection{Motivation}
Natural and social disasters, such as fires, earthquakes, landslides, and flooding, are the most influential events affecting human societies. Moreover, it is difficult to predict the occurrence of natural disasters and to minimize the damage resulted from them. Many countries have suffered from human and financial losses due to unexpected fires and earthquakes, and thousands of people lose their lives and property every year\cite{Adu-Manu2018}. Gangneung (Republic of Korea); California (USA); and New South Wales (Australia), are a few examples of such places that have endured major damage from large forest fires in 2019. In addition, large earthquakes occurred in California (USA), July 2019, and Yamakata (Japan), June 2019, causing damage on a national scale. The best way to minimize the damage caused by disasters is to predict them early and respond to them proactively. In case of a fire, it is difficult to extinguish when it grows to a certain scale; this is why it is important to recognize, in a timely manner, the possibility of a fire occurring and to suppress its growth. In case of an earthquake, it is important to predict an earthquake by identifying its likelihood and responding swiftly to disasters in close proximity to the epicenter. Early detection of disaster occurrences extends the golden period, ensuring safe evacuation in a calamitous situation\cite{Qian2018}.

To design a permanent disaster monitoring system, we have to overcome three constraints. One constraint is that sensor nodes have limited energy sources\cite{Chettri2020}. Minimizing the energy consumption is important to expend a lifetime of sensor nodes. It reduces the lifetime of sensor nodes to collect and transmit data under the limited energy source in the wild environment, such as on a mountain or underground, where no electric power infrastructure is installed\cite{Militano2016}. To tackle this issue, installing power grids in forests and underground leads to environmental degradation, as well as incur colossal construction costs. Dead sensor nodes contribute to environmental pollution due to their chemical components. Energy harvesting is an alternative to power grids. Another constraint is the instability of harvesting energy\cite{Deng2019}. Since harvesting methods have a large variance of supplied power, an algorithm that manages the power variance is necessary\cite{Lee2017a}. We suggest a method to use harvesting sources as a stable power line considering the characteristics of a harvester. The other constraint is the complexity of energy conservation algorithms. Complexity for operating an algorithm causes larger energy consumption. High complexity makes it hard for real-time operations.

To overcome all constraints, we propose a novel energy management protocol and three energy conservation algorithms. This study proposes ASMP for the most efficient disaster monitoring protocol. The contributions of this study are summarized below:
\begin{itemize}
\item {ASMP systematically manages all devices in the network in order to ensure three requirements, self-sustainability, QoS, and low complexity. ASMP makes it possible for sensor nodes to activate three proposed algorithms, ASA-m, CASA, and RASA, autonomously by considering deployed power environment.}
\item {ASMP recharges the energy of sensor nodes to guarantee self-sustainability by using RASA. It makes energy consumption smaller than Exponentially Weighted Moving Average (EWMA) of harvested energy when sensor nodes in the energy-hungry state.}
\item {We define the target data for QoS optimized operation. It means the particular data that a sensor node has to gather for satisfying spatial resolution on a server-side. From this definition, we define the coordinates plane that is in charge of data mapping.}
\item {To ensure QoS requirements, ASMP utilizes both sampled and predicted data. CASA and RASA cause sampling blanks for reducing energy. The server fills in sampling blanks by using predicted data at the ASA-m sampling point.}
\item {We develop a simple estimation method to reduce calculation complexity. ASA-m calculates the slope of two sampled points. We define it as the mean velocity of the data transition. ASA-m estimates the sampling period by using the trend of mean velocity.}
\end{itemize} 
Section \textbf{II} presents related research on energy conservation methods and introduces energy harvesting techniques. Section \textbf{III} introduces the requirements to organize the endless monitoring system. Section \textbf{IV} explains the details of the proposed algorithms. Section \textbf{V} includes mathematical definitions of energy consumption of sensor nodes. Section \textbf{VI} explains our design methods to organize efficient monitoring systems, as well as the simulation environments. Section\textbf{VII} shows simulation results and the performance of our system. Section \textbf{VIII} concludes the paper.

\begin{figure*}[!ht]
\setlength{\fboxsep}{0pt}%
\setlength{\fboxrule}{0pt}%
\begin{center}
\includegraphics[scale=0.55]{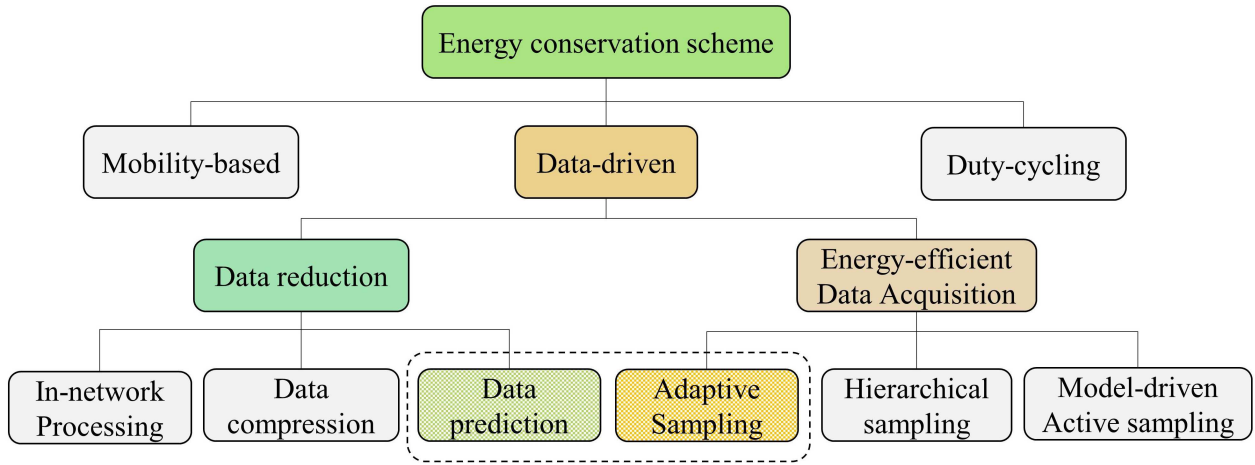}
\end{center}
\caption{The chart of energy conservation research sections.\label{F1}}
\end{figure*}

\section{Diverse approaches for a self-sustainable monitoring system}
There have been a variety of approaches to meet the energy constraint of sensor nodes in energy harvesting wireless sensor networks (EH-WSNs). Most of these approaches are divided into two main parts; energy management and energy harvesting. These approaches focus on optimization of the consumed energy. Energy harvesting has diverse techniques, including solar panel, wind, and radio frequency (RF) transfer. Wherein many of them can be combined.

\subsection{Energy Management Scheme}
There are diverse methods to reduce the energy consumption of sensor nodes and to optimize their operation. Khan \textit{et al.}\cite{Khan2015} and Anastasi \textit{et al.} \cite{Anastasi2009} introduce definitions and survey of energy management or conservation in EH-WSNs. Until now, energy constraint problems still remains unsolved. Figure~\ref{F1}\footnote{This figure was modified and redrawn from Figure 3 of Anastasi \textit{et al.}\cite{Anastasi2009}.} shows the detailed assortment of the energy conservation scheme\cite{Guo2019}. Energy conservation schemes are categorized into three sub-parts; mobility-based, data-driven, and duty cycling\cite{Yick2008}. 

\subsubsection{Mobility-based}
The mobility-based is a part to reduce the communication energy. Using mobile syncs or relay nodes, sensor nodes decrease the transmission distance and reduce energy consumption caused by multi-hop transmission. Mobile syncs or relay nodes collect the sampled data by moving around the monitoring area along an optimized path. From this perspective, the moving path of mobile syncs is the most important factor because mobile syncs should periodically and rapidly gather all the sampled data in the monitored area. The mobility-based part proposes optimized paths for mobile sync nodes, as in SCAN, HILBERT, and DREAMS\cite{In2006,HALDER201682}.

\subsubsection{Duty-cycling}
This is to improve network lifetime through the use of the network topology control, sleep and wake-up, and low duty-cycle MAC protocols\cite{Khan2015}. When sensor nodes have no role in their network, they are switched to a sleep mode and wait to be put in charge of some demands. The advantage of duty cycling is in minimizing the energy consumption by changing status, such as sleep and wake-up and duty-cycle MAC protocol\cite{Lai2010}. 
 
\subsubsection{Data-driven}
This research area focuses to reduce energy consumption for data acquisition\cite{Elshrkawey2018}. Sensor nodes try to reduce the sampling operations by using data prediction or efficient data acquisition algorithms, such as ASAs and low energy adaptive clustering hierarchy (LEACH)\cite{Handy2002}.

{In adaptive sampling, Alippi \textit{et al.}\cite{5208303} introduced ASA. It is focused to fit the sampling rate to the variation period of the monitored environment using the Nyquist sampling theorem. Srbinovski \textit{et al.}\cite{Srbinovski2016} proposed energy aware ASA to improve the lifetime of sensor nodes by applying their energy variables. It adjusts the sampling rate considering remained battery. Due to the degrading input energy variables, QoS decreases consecutively. Shu \textit{et al.}\cite{Shu2017} indicated the drawbacks of previous ASAs and proposed a data-driven ASA based on the absolute difference between current and average values of last $N$ numbers of sampled data. It adjusts the sampling rate by multiplying a sigmoid function among $[0,2]$. Gindullina \textit{et al.}\cite{gindullina2020energy} proposed an energy aware model, combining a sigmoid function in\cite{Shu2017} and energy variables. For ensuring lifetime and QoS in real-world WSNs, we propose a novel protocol based on ASAs. It recharges energy-hungry sensor nodes.}

{In dual prediction\cite{8664582}, Tan \textit{et al.}\cite{Tan2016} proposed an adaptive filter based on a spatial-temporal correlation. It is hard to apply in low-performance sensor nodes due to high complexity. Raza \textit{et al.}\cite{Raza2015} and Barton \textit{et al.}\cite{Barton2016} introduced an less complex adaptive filter for short-term linear behavior. It is improper for disasters having non-linear behavior. Wang \textit{et al.}\cite{Wang2007} proposed a filter to trace polynomial behavior as a non-linear model. To reduce model complexity and trace disasters, we propose a novel adaptive filter including the variability of environments. It is advantageous both linear and non-linear behavior.}

{Our proposed protocol combines two parts (i.e., Adaptive sampling and Data prediction) for both guaranteeing self-sustainability and minimizing sensing data. Using data prediction with trends, proposed algorithm expects the time to occur the certain amount of environment change, called a meaningful change. This time is to be a sampling period.} Problem statements and requirements for the permanent monitoring system are as follows:
\begin{itemize}
\item{\textbf{QoS satisfaction:} Disaster monitoring systems have a strict QoS requirement. All algorithms have to satisfy the spatial resolution requirements for the reliable disaster monitoring system. They have to trace all dynamic environment transitions.}
\item{\textbf{Self-sustainability:} Using only optimized approaches is not enough to satisfy the self-sustainability of sensor nodes. The energy conservation algorithm has to make energy consumption less than EWMA of harvest in the energy-hungry state.}
\item{\textbf{Low complexity:} Heavy load algorithms cause larger energy consumption. It is difficult to operate them in real-time on low-performance sensor nodes. WSNs require low complex and effective data gathering algorithms.}
\end{itemize}

\begin{table*}[!t]
\small\sf\centering
\renewcommand{\arraystretch}{1.5}
\renewcommand{\multirowsetup}{\centering}
\caption{Comparison of diverse energy harvesting methods and examples\label{T1}}
\begin{tabular}{>{\centering}m{2cm}>{\centering}m{2cm}>{\centering}m{2.5cm}>{\centering}m{2cm}>{\centering}m{2cm}>{\centering}m{2.5cm}>{\centering}m{2cm}}
\hline
Energy Harvesting Sources& Conversion Efficiency ($\%$)&Amount of energy Harvested ($mW/cm^2$)& Example \par Products& Output Voltage($V$)& Maximum Output Power ($mW$) & Dimensions ($mm^2$) \tabularnewline
\hline
\multirow{3}{2cm}{Solar\cite{Sudevalayam2011}$^,$\cite{Rand2007}} & \multirow{3}{2cm}{CdTe : $28\%$, Si : $15\sim24\%$, CIGS : $20\%$} & \multirow{3}{2.5cm}{$10.5\sim25.6$} & SC10050 & $5$ & $500$ & $75\times60$ \tabularnewline
\cline{4-7}
 &&&SZGD8855&$3.84$&$550$&$88\times55$ \tabularnewline
 \cline{4-7}
 &&&DIYJOA solar&$6$& $2\times10^{3}$ &$109.5\times131.5$\tabularnewline
\hline
\multirow{2}{2cm}{Wind\cite{Qian2018}}&\multirow{2}{2cm}{$35-45\%$} &\multirow{2}{2.5cm}{-} & NE-100S2 & $12/24$ & $10\times10^{4}$ & $1150$ \par \scriptsize{(Wheel diameter)} \tabularnewline
\cline{4-7}
&&& EW400 & $24$ & $40\times10^{4}$ & $1550$ \par \scriptsize{(Wheel diameter)} \tabularnewline
\hline
RF Transfer\cite{Lu2014}\cite{Ren2018} & $0.4\%${\scriptsize($-40dBm$)}\par $18.2\%${\scriptsize($-20dBm$)}\par \scriptsize{input power} & - & P2110-EVB & $2.0\sim5.5$ & $275$ & $13.5\times14$ {\scriptsize(Without antenna)} \tabularnewline
\hline
Piezoelectric\cite{Priya2017} & - & $1.51$& S129-H5FR-1803YB & $5\sim35$ & $19.6$ & $55.4\times23.4$ \tabularnewline
\hline
Thermal\cite{Kishore2018} & $1-10\%$ & $ 4.9\times10^{3}$ & CP85138 & $2.1$ & $11.1\times10^{3}$ & $15\times15$\tabularnewline
\hline
\end{tabular}\\[2pt]
\centering\footnotesize Please refer to the datasheets for various modules\\[3pt]
\end{table*}

\subsection{Energy Harvesting Techniques}
Because of the limitation of battery capacity, the sensor nodes in EH-WSN need to obtain energy from their environment independently through harvesting techniques or other methods\cite{Tang2018}. Sudevalayam \textit{et al.}\cite{Sudevalayam2011} and Adu-Manu \textit{et al.}\cite{Adu-Manu2018} introduce definitions and survey of energy harvesting techniques. Table~\ref{T1} summarizes information and product examples of energy harvesting technologies. The solar is divided into two systems, namely, thermal conversion and photo-voltaic conversion\cite{Rand2007}. Photo-volatic converts the photon energy of light to electricity by using semiconductor material. It is advantageous to be utilized in sensor nodes due to a simple structure and small size. Thermal conversion entails a method to heat a medium using the thermal energy of the sun. This study does not assume thermal conversion due to its complex structure. It is not suitable for the tiny sensor nodes of EH-WSNs\cite{Lee2013a}. Wind energy harvesting techniques use the flow of air. Wind harvesting has a great conversion efficiency, but it is hard to find small-sized harvesters\cite{Qian2018}. Simultaneous Wireless Information and Power Transfer (SWIPT) and RF energy transfer are vital considerations. RF energy transfer has some benefits since it controls amount of harvested power\cite{Lu2014,Ren2018}. Nonetheless, the difficulty in transmitting over long distances still exist. Furthermore, high-powered electromagnetic waves affect the human body. Harvesting energy is an unstable power supplement because it is sensitive to the change of environmental factors, such as climate and weather. The energy conservation algorithm should be designed to adapt to large power changes.

\section{The system model and requirements for an endless monitoring system}
A set of sensor nodes ($\mathbf{X}=\{X_1,X_2,\cdots,X_i\}$) are placed in the forest. They monitor the temperature in the forest. The fire monitoring system is organized by a server and three types of sensor nodes, End, Relay, and Sync. The monitoring signal at the place located sensor node $X$ is defined as $\mathbf{S}(t)$. The server and sensor nodes in the system have a set of target data ($\mathbf{B}=\{B_1,B_2,\cdots,B_{n-1},B_n,\cdots,B_{max}\}$), based on the QoS requirement of monitoring system. We define a ($t$, $\mathbf{B}$) coordinates plane as a data plane. The sensor node $X$ calculates the sampling period, which is the time required for the meaningful change ($B^{'}$). The meaningful change ($B^{'}$) means the gap between the current and next target data. It is y-axis grid size, i.e., a spatial resolution on a data plane. It follows the spatial difference method of statistics at the plane. It means the difference between two points of the plane\cite{cressie2015statistics}. It is the same as the QoS identifier. $B^{'}$ is a fixed real number or variable and it is set by the system requirement. The sensor node $X$ gathers the data ($D_n=\mathbf{S}(\sum_{\beta=1}^{n}T_s^{\beta})+\mathbf{N}[n]$) at the every sampling periods. $\mathbf{N}[n]$ denotes sensor noise. The next sampling period ($T_s^{n+1}$) is optimized in the transition of the monitored environment. A set of sampled data is defined as $\mathbf{D}=\{D_1,D_2,\cdots,D_{n-1},D_n,\cdots\}$. ${}^{\forall}D_n$ are posited to data plane. $D_n^{event}$ means the criterion value for disaster alarm. The system operates the warning process if the sampled data is over the $D_n^{event}$. The solar panel for energy harvesting is embedded in sensor node $X$. Sensor node $X$ is in the sleep state during a remained sampling period except for sampling operation time because the proposed system uses the sleep and wake-up protocol. Sensor node $X$ senses the residual ($E_r$) and harvesting ($E_h$) energy state by itself. The detailed definitions for energy modeling are in Section \textbf{V}. 

Sensor node $X$ has to satisfy the requirements as follows. First, it periodically gathers the environmental data to detect symptoms of a disaster early whereas guaranteeing a stable battery state or lifetime. Second, sensor node $X$ simultaneously conducts sampling and transmitting operations for a real-time monitoring and more accurate disaster detection\cite{Srbinovski2016}. The disaster detection of a single sensor node causes high false and miss alarm probability because it is disabled double-checking. A few false and miss alarms diminish the reliability of the disaster-monitoring system. To reduce unnecessary confusion and damage from false alarms, the server has to decide after synthetically considering sufficient data from all sensor nodes. Furthermore, sensor node $X$ detects and transmits only the target data of the system to reduce the transmission delay. Reducing data traffic is helpful to minimize the transmission delay by decreasing data collision and re-transmission. Finally, sensor node $X$ takes on one of the different roles in the network by time, such as an end, relay, and sync since energy fairness is also an important requirement for network lifetime. The monitoring system organizes self-directly, according to the changing of the network environment.

\section{Adaptive Sensor node Management Protocol}
\begin{figure}[!t]
\setlength{\fboxsep}{0pt}
\setlength{\fboxrule}{0pt}
\centerline{\includegraphics[width=8.8cm]{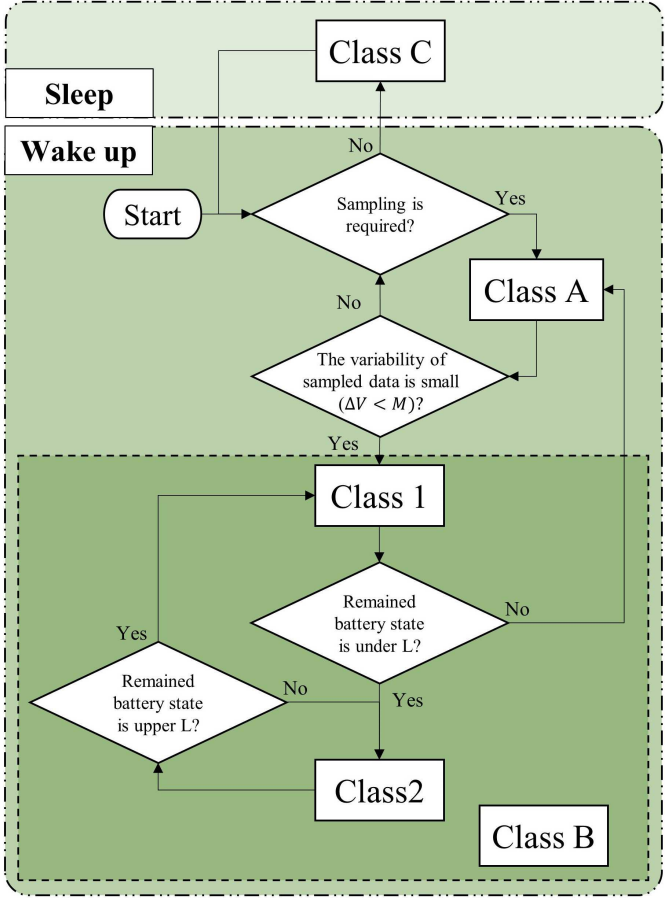}}
\caption{The flowchart of a sensor node applied ASMP.\label{F6}}
\end{figure}

ASMP is the most effective network protocol for permanent disaster monitoring. The proposed protocol satisfies all the requirements earlier introduced in Section \textbf{II}. ASMP has operational priority in the order of self-sustainability and QoS requirements. ASMP tries to collect enough data satisfying QoS under energy constraints. Sensor nodes self-directly control the sampling duty to save energy. A server supplements QoS requirements. ASMP classifies sensor nodes into three classes and two sub-classes. Figure~\ref{F6} shows the operation flow in all sensor nodes of ASMP.

Class A in ASMP is the initial state of sensor nodes. Sensor nodes in Class A gather only target data by utilizing the sampling rate of ASA-m (Section \textbf{IV-A}). Unlike fixed sampling, ASA-m minimizes the wasted power consumption caused by unnecessary data gathering. ASA-m gives an optimized sampling rate based on a spatial resolution. Class A includes the sensor nodes that have a great battery state.

Class B is the state that the battery of sensor nodes significantly decreases despite of the adoption of ASA-m. Sensor nodes adopt two additional energy conservation algorithms, called CASA and RASA. Class B classifies sensor nodes into two sub-classes, Class 1 and Class 2. Class 1 means the group of sensor nodes using CASA. Class 2 denotes the group of sensor nodes using RASA. CASA has a greater energy conservation performance than ASA-m, while maintaining a slightly higher error rate. The system server replaces the next sampling data by a predicted. The sensor node takes a rest for a sampling period when the monitoring environment is linearly changing. {CASA factor ($\textbf{M}$) is a criterion to judge the linearity of environment change based on the variability of data in Equation~(\ref{casa}).} More details of CASA are described in Section \textbf{IV-B}. 

ASMP classifies sensor nodes into Class 2 when the energy state is under a certain energy threshold ($L$). $L$ means an emergency energy level of sensor nodes. If harvested energy goes to $0$ at night, the rest of lifetime ($T_{sus}$) of sensor nodes in (\ref{eq:4_f})-(\ref{eq:4}) noticeably decreases. The lifetime ($T_{sus}$) has to be always larger than the maximum interval of poor or zero harvest time ($\mathbf{I_o}$). Based on Equation~(\ref{L}), ASMP sets the energy threshold ($L$) as a parameter in (\ref{L_r}) or a preset constant.
\begin{IEEEeqnarray}{c} \label{L}
\frac{L}{O_mf_{s}E_{sp}}=T_{sus} \geq max(\mathbf{I_o})\\
L \geq max(\mathbf{I_o})O_mf_sE_{sp} \label{L_r}
\end{IEEEeqnarray}
{where $O_m$ is operation modes of a sensor node. It has one of values, 1 or 2 or $C$ in (\ref{eq:4_f})-(\ref{eq:4}). $E_{sp}$ is the sum of energy consumed by sampling at once. $f_s$ denotes a sampling rate.}

Under $L$ state, sensor nodes activate RASA. They compare consumed and harvested energy and reduce the sampling rate until the consuming energy is less than the harvesting that. The RASA factor ($\textbf{N}$) in Equation~(\ref{rasa_f_f}) is a multiplier that extend a sampling period of sensor nodes. Through this method, the depleted sensor nodes have a chance to recover their energy state. RASA is explained in more detail in Section \textbf{IV-C}.

Class C denotes the sleep state of sensor nodes. If a sensor node does not have any roles in the network, ASMP puts it into Class C, and the sensor node is returned to Class A when it is needed. 

\begin{figure}[t]
\setlength{\fboxsep}{0pt}
\setlength{\fboxrule}{0pt}
\centerline{\includegraphics[width=8.8cm]{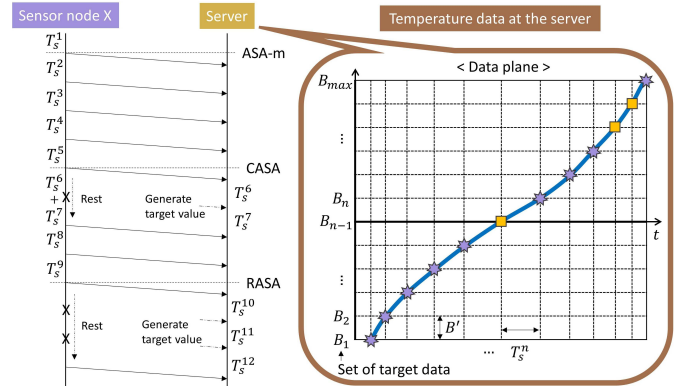}}
\caption{The example of data acquiring operation at a server-side using the ASMP protocol ($\star$=sampled data, $\square$=estimated data).\label{exap}}
\end{figure}

Figure~\ref{exap} shows an example of ASMP operation and a data plane on server-side. {During the time ($T_s^1\sim T_s^4$), ($T_s^5\sim T_s^8$), and ($T_s^{9}\sim T_s^{12}$), it shows a ASA-m, CASA, and RASA operation.} After the time ($T_s^5$), the sensor node shares its own class information to the server. The server indicates that a next sampled data will not arrive after the ASA-m sampling period ($T_s^6$). Then, the server generates predicted data after the sampling period ($T_s^6$) in order to inhibit the degrade service quality. Through this process, the monitoring system fills in blanks of sampling and guarantees QoS requirements. 

{We summarize three main contributions of ASMP. Firstly, ASMP guarantees both a self-sustainability of sensor nodes and a spatial resolution of disaster monitoring application. Secondly, ASMP reduces and disperses network traffics since sensor nodes have all different sampling and transmitting periods. It reduces network problems, such as collision, re-transmission and delay.} Lastly, ASMP recognizes network performances and supports QoS requirements. The server knows when the next traffic should arrive since ASMP calculates sampling time intervals not only on a sensor node-side but also on a server-side. Using this property, ASMP gives resilience to changes in network performance such as delay, jitter, and packet loss probability.

\subsection{Adaptive Sampling Algorithm for monitoring (ASA-m)}
\begin{figure}[!t]
\setlength{\fboxsep}{0pt}%
\setlength{\fboxrule}{0pt}%
\centerline{\includegraphics[width=8.8cm]{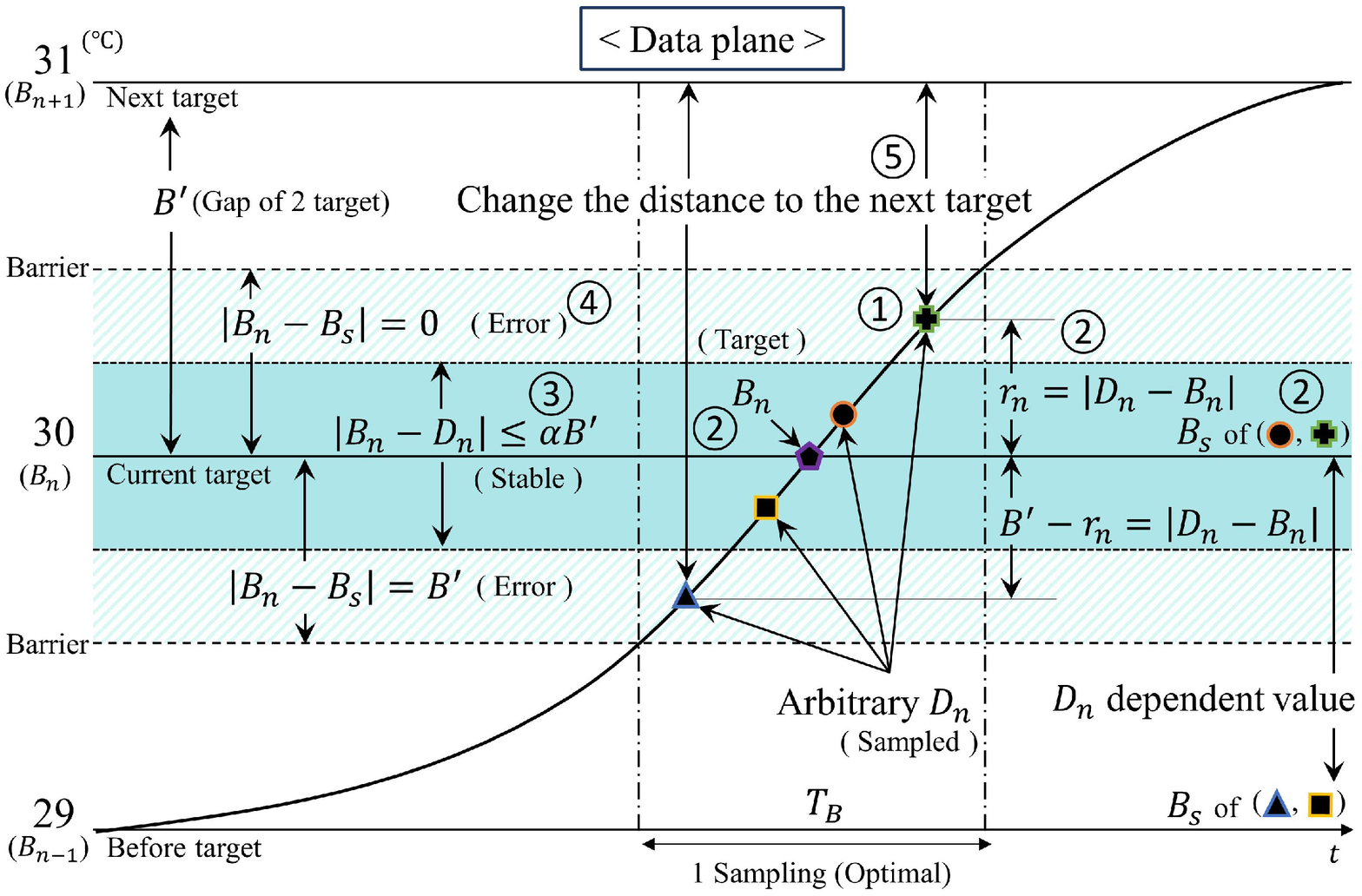}}
\caption{{The data plane and target distance adjustment procedures, where $B_n$ is $30\,^{\circ}\mathrm{C}$. $D_n$ is $\mathbf{+}$ of marks, $\bigtriangleup$, $\Box$, $\bigcirc$, and $\mathbf{+}$. Notations refer to Table~\ref{notation_ASA}.}\label{F15}}
\end{figure}

{The sampling rate of previous ASAs continuously decreases according to the degraded energy state of sensor nodes. It causes QoS problems. Furthermore, previous ASAs misunderstand a sampling noise of a sensor as the changing of data, which causes unnecessary energy consumption and increases network congestion. To overcome these limitations, we propose an ASA-m considering a meaningful change and QoS requirements. ASA-m focuses on detecting only meaningful changes requiring by a monitoring system. ASA-m tries to sample the next target data, which is one of target data set ($\mathbf{B}$). ASA-m estimates the next period for sampling the next target data considering the properties for environmental change. Nature environment changes are continuous and follow an energy transfer mechanism, such as heat transfer.} To trace this change, ASA-m calculates optimized sampling rate by using a proposed mean velocity based prediction in Section \textbf{IV-A-}$2)$. QoS optimum sampling rate ($f_s$) in (\ref{F_asa}) is to sample at once within the range ($T_B$) between two barriers in Figure~\ref{F15}.
\begin{IEEEeqnarray}{c}
f_{s}= \frac{1}{T_{B}}
\label{F_asa}
\end{IEEEeqnarray}

\begin{figure}[!t]
\setlength{\fboxsep}{0pt}
\setlength{\fboxrule}{0pt}
\centerline{\includegraphics[width=8.8cm]{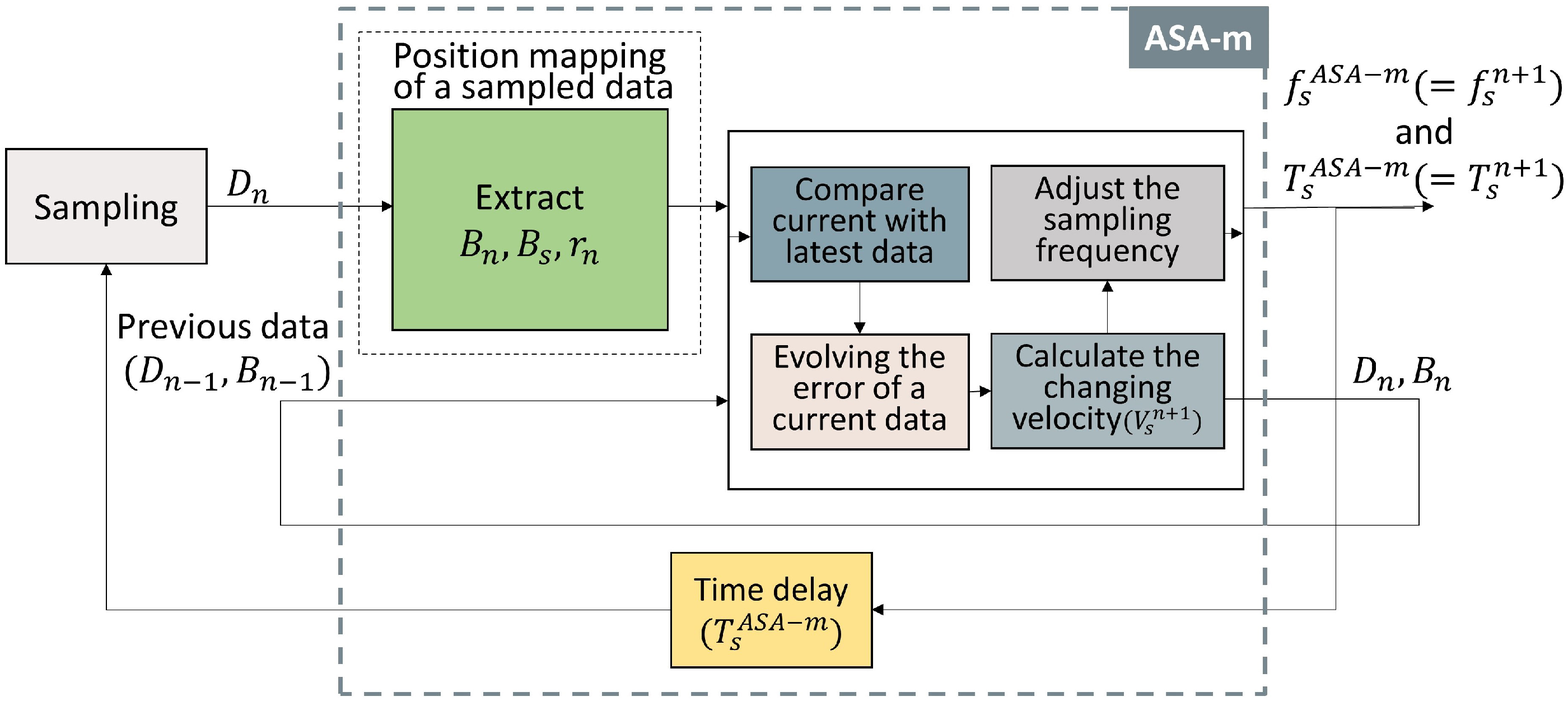}}
\caption{The block diagram of ASA-m. {Notations refer to Table~\ref{notation_ASA}.}\label{diagram}}
\end{figure}

To calculate an optimized sampling period, ASA-m uses three-sampled data (i.e., a current and two latest data). Figure~\ref{diagram} provides entire processes to calculate ASA-m sampling rate ($f_s^{ASA-m}$). Firstly, ASA-m places all sampled data in the data plane. When a sensor node samples a data ($D_n$), the sensor node rounds off the sampled data to the nearest target value ($B_n$). Additionally, ASA-m computes the quotient ($B_s$) and the remainder ($r_n$) of $D_n/B^{'}$ to minimize sampling errors, where $B^{'}$ is the size of meaningful change. Using these values, a sensor node positions sampled data to the data plane ($t, \textbf{B}$) as in Figure~\ref{F15}. {The relation between the next target data ($B_{n+1}$) and sampling period ($T_s^{n+1}$) defines as $B_{n+1}=B_n\pm B^{'}=D_n+V_c^{n+1}T_s^{n+1}$, where $D_n$ is a sampled data, $V_c^{n+1}$ is a predicted velocity in (\ref{V_n+1}). ASA-m predicts $V_c^{n+1}$ through MVP (Section \textbf{IV-A-}$2)$) and next derives $T_s^{n+1}=|B_n\pm B^{'}-D_n|/V_c^{n+1}$, where $|B_n\pm B^{'}-D_n|$ means the distance to target.} Based on these definitions, ASA-m adjusts a sampling rate to minimize errors by using two types of distance in Section \textbf{IV-A-}$1)$.


\subsubsection{{The two types of distance analysis}}
ASA-m compares current and previous data. ASA-m evaluates the sampling error using two types of distance: $|B_n-B_{n-1}|$ and $\vert D_{n}-B_{n}\vert$. 

{$|B_n-B_{n-1}|$ is the distance between two sampled data. $|B_n-B_{n-1}|>B^{'}$ means that an estimated sampling period misses some target data. ASA-m multiplies as much as the count of missed target data to the next sampling rate. Otherwise, if $|B_n-B_{n-1}|=0$, ASA-m sufficiently extends the next sampling period to collect next target data ($B_{n+2}$). The server generates target data ($B_{n+1}$) after the time ($T_s^{post}$).}

$\vert D_{n}-B_{n}\vert$ denotes the distance between a sampled and target values. It follows a temporal difference method of statistics\cite{sutton1988learning}. {Figure~\ref{F15} shows a data plane and target distance adjustment procedure. $\alpha$ denotes a parameter to control the sensitivity of ASA-m. It adjusts the scope of stable.} $\vert B_{n}-B_{s}\vert=0$ means that $D_{n}$ exists on the upper side of the target value line. {ASA-m reduces the distance to the next target data when it is $\vert D_{n}-B_{n}\vert>\alpha B^{'}$ in (\ref{D_B}).} $\vert B_{n}-B_{s}\vert=B^{'}$ means that $D_{n}$ locates under of the target value line. {ASA-m extends the distance to the next target when it is in $\vert D_{n}-B_{n}\vert>\alpha B^{'}$.} Using this approach, sensor nodes adjust the sampling rate to improve sampling accuracy. 

After computing the distance to the next target data, the next step is to predict the next mean velocity ($V_c^{n+1}$) by using a proposed prediction model (Section \textbf{IV-A-}$2)$) including the variability of environment. Sensor nodes estimate the next optimal sampling period ($T_{s}^{n+1}$), e.g., when sampling is stable ($\vert D_{n}-B_{n}\vert < \alpha B^{'}$), $T_{s}^{n+1}$ is $B^{'}/|V_c^{n+1}|$ as in (\ref{T_n+1}).

\begin{figure}[!t]
\setlength{\fboxsep}{0pt}%
\setlength{\fboxrule}{0pt}%
\centerline{\includegraphics[width=8.8cm]{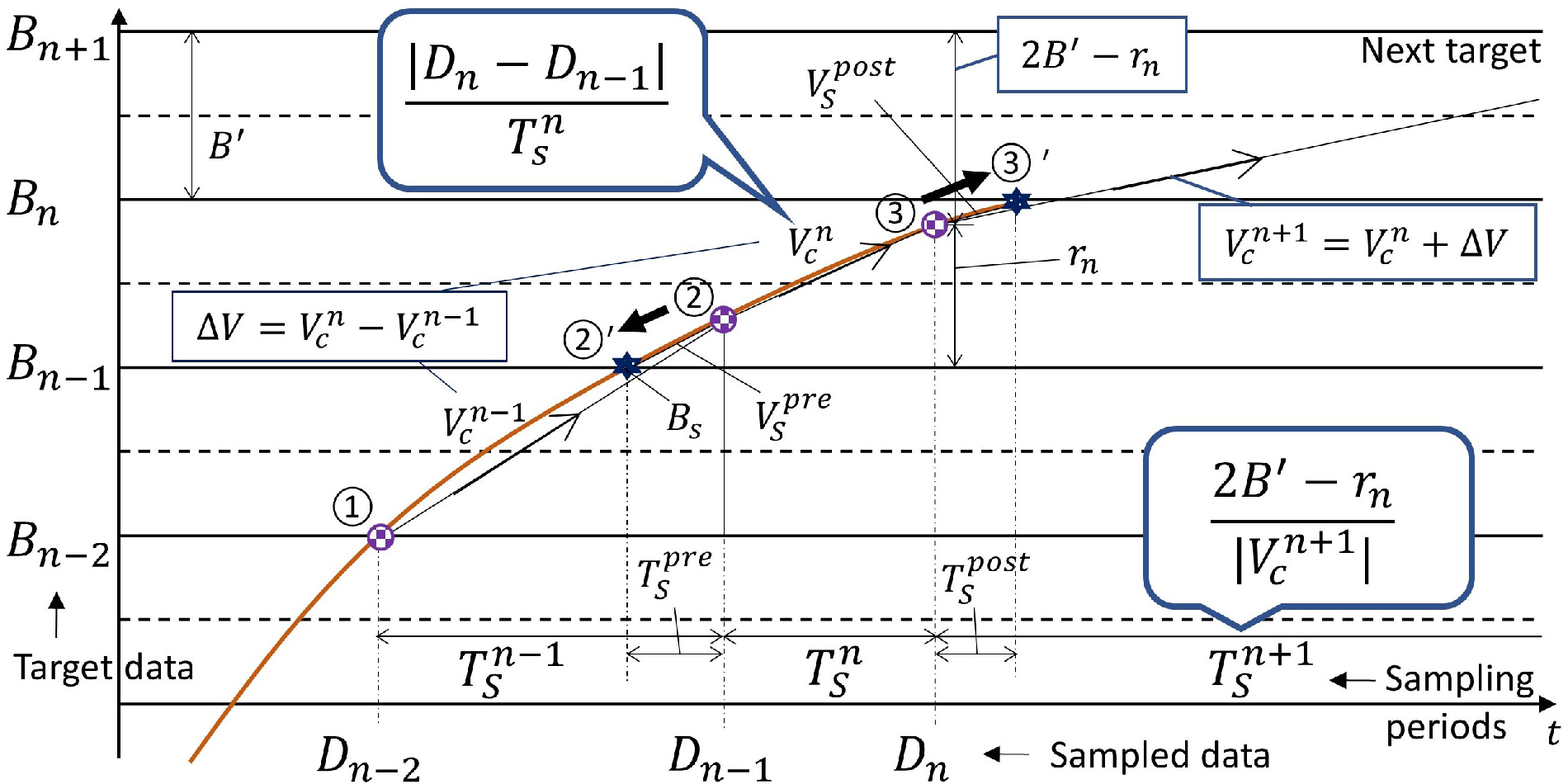}}
\caption{{The next sampling period estimation process based on MVP. It represents for mathematical definitions and notations in Table~\ref{notation_ASA}.}\label{F16}}
\end{figure}

\begin{table}[!t]
\renewcommand{\arraystretch}{1.3}
\caption{Notations}
\label{notation_ASA}
\centering
% Some packages, such as MDW tools, offer better commands for making tables
% than the plain LaTeX2e tabular which is used here.
\begin{tabular}{m{1.5cm}m{6.5cm}}
\hline
$\boldsymbol{Symbol}$ & $\boldsymbol{Description}$\\
\hline
$T_{s}^n$&\text{The $n$-th sampling period}\tabularnewline
$D_{n}$&\text{The $n$-th sampled data}\tabularnewline
$B_{n}$&\text{The nearest value of target set from $n$-th sampled data}\tabularnewline
$B^{'}$&\text{The meaningful change derived by a spatial resolution}\tabularnewline
$B_{s}$&\text{The target data just below the $n$-th sampled data}\tabularnewline
$r_{n}$&\text{The distance between $D_{n}$ and $B_{s}$}\tabularnewline
$V_{c}^{n}$&\text{The mean velocity of $n$-th data}\tabularnewline
$\alpha$&\text{The sensitivity tuner}\tabularnewline
\hline
\end{tabular}
\end{table}

\begin{table}[!t]
\label{table_asa}
\centering
% Some packages, such as MDW tools, offer better commands for making tables
% than the plain LaTeX2e tabular which is used here.
\begin{tabular}{m{8.3cm}}
\hline
$\mathbf{Algorithm\,1:}$ Adaptive\,sampling\,algorithm\,for\, monitoring\, (ASA-m)\\
\hline
$\mathbf{Input\,:}$ Sampled data ($D_{n}$); \par
$\alpha=const$,\,$B^{'}=const$,\,$f_{s}^{n+1}=initial$ \\[3pt]
$\mathbf{Output\,:}$ Adaptive sampling rate ($f_{s}^{ASA-m}$) \par\\[3pt]
$\mathbf{Sensor\,node\,process:}$ \par
$\mathbf{While(true)}$ \,\{ \par
\quad $B_{n}\leftarrow$ $B^{'}*round off(D_{n}/B^{'})$,\,$B_{s}\leftarrow B^{'}*round down(D_{n}/B^{'})$\par
\quad $r_{n}\leftarrow remains(D_{n}/B^{'})$,\,$V_{c}^{n}\leftarrow (D_{n}-D_{n-1})*f_{s}^{n}$ (in~(\ref{V_c_f}));\par
\quad $V_{c}^{n+1}\leftarrow V_{c}^{n}+(V_{c}^{n}-V_{c}^{n-1})$ (in~(\ref{V_n+1}))\par
\par \\[3pt]
\quad $\mathbf{Case\,I}$ The increasing trend ($D_{n}>D_{n-1}$) (in~(\ref{T_n+1}))\par
\quad\quad $\mathbf{if}$ missed some target ($\vert B_{n}-B_{n-1}\vert>B^{'}$)\par
\quad\quad\quad The number of missed target ($q_{over})\leftarrow \vert B_{n}-B_{n-1}\vert/B^{'}$;\par
\quad\quad\quad $f_{s}^{n+1}= q_{over}\times f_{s}^{n}$; \par
\quad\quad $\mathbf{else\,if}$ sampling is stable ($\vert D_{n}-B_{n}\vert < \alpha B^{'}$) \par
\quad\quad\quad $f_{s}^{n+1}= \vert V_{c}^{n+1}\vert/B^{'}$;\par
\quad\quad $\mathbf{else}$ sampling has some error \par
\quad\quad\quad $f_{s}^{n+1}= \vert V_{c}^{n+1}\vert/(2B^{'}-r_{n})$;\par
\quad\quad\quad $B_{n}=round(B_{n}+(V_{c}^{n+1}*T_{s}^{post})$);\par
\par \\[3pt]
\quad $\mathbf{Case\,II}$ The decreasing trend ($D_{n}<D_{n-1}$) (in~(\ref{T_n+1}))\par
\quad\quad $\mathbf{if}$ missed some target ($\vert B_{n}-B_{n-1}\vert>B^{'}$) \par
\quad\quad\quad The number of missed target ($q_{over})\leftarrow \vert B_{n}-B_{n-1}\vert/B^{'}$;\par
\quad\quad\quad $f_{s}^{n+1}= q_{over}\times f_{s}^{n}$; \par
\quad\quad $\mathbf{else\,if}$ sampling is stable ($\vert D_{n}-B_{n}\vert < \alpha B^{'}$) \par
\quad\quad\quad $f_{s}^{n+1}= \vert V_{c}^{n+1}\vert/B^{'}$;\par
\quad\quad $\mathbf{else}$ sampling has some error\par
\quad\quad\quad $f_{s}^{n+1}= \vert V_{c}^{n+1}\vert/(B^{'}+r_{n})$;\par
\quad\quad\quad $B_{n}=round(B_{n}+(V_{c}^{n+1}*T_{s}^{post})$);\par
\par \\[3pt]
\quad $\mathbf{Limit}$ a sampling rate as in~(\ref{T_a})\par
\quad\quad $\mathbf{if}$ ($f_{s}^{n+1} < f_{s}^{max}$) \par
\quad\quad\quad $f_{s}^{n+1}=f_{s}^{max}$; \par
\par \\[3pt]
\quad $\mathbf{Feedback\,Inputs}$ \par
\quad\quad $D_{n-1}\leftarrow D_{n}$,\, $B_{n-1}\leftarrow B_{n}$,\, $r_{n-1}\leftarrow r_{n}$,\,$V_{c}^{n-1}\leftarrow V_{c}^{n}$,\par
\quad\quad $f_{s}^{n}\leftarrow f_{s}^{n+1}$\par
\par\\[3pt]
$\mathbf{Server\,process:}$ Generate a predicted data after $T_{s}^{post}$ in~(\ref{T_post_I})-(\ref{T_post_D})\\
$\mathbf{if}$ sampling has some error ($\alpha B^{'} < \vert D_{n}-B_{n}\vert < B^{'}$)\\
\quad $D_{pre}=\left(D_{n}+(V_{c}^{post}*T_{s}^{post})\right)$; \\
\hline
\end{tabular}
\end{table}
\begin{figure*}[!t]
\setlength{\fboxsep}{0pt}
\setlength{\fboxrule}{0pt}
\centerline{\includegraphics[width=18.1cm]{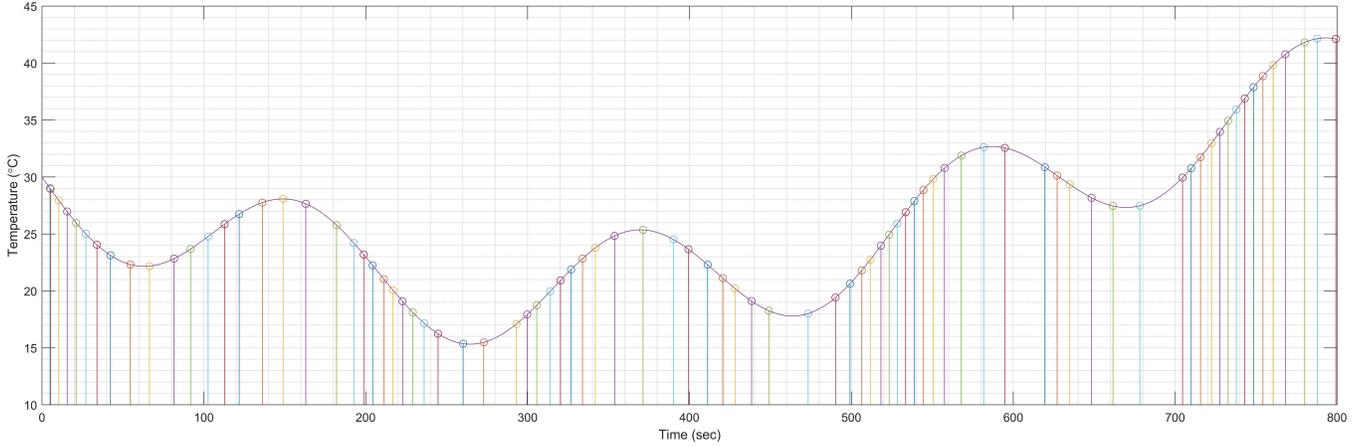}}
\caption{The sampling result of ASA-m using a dynamic transition data set, when $\alpha$ is 0.2 and $B^{'}$ is 1 Celsius.\label{F17}}
\end{figure*}

\begin{figure}[!t]
\setlength{\fboxsep}{0pt}%
\setlength{\fboxrule}{0pt}%
\centerline{\includegraphics[width=8.8cm]{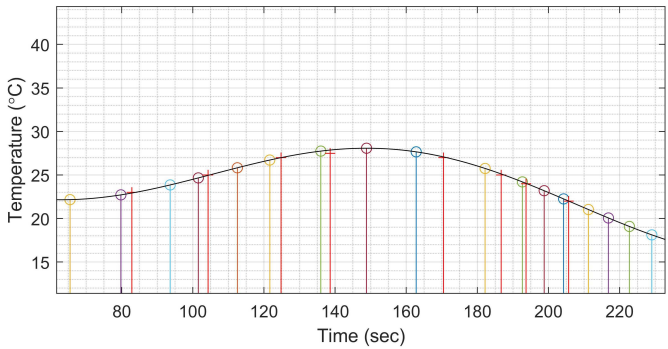}}
\caption{The data collection result of a network server using ASA-m, Mark ($\bigcirc$) means the sampled data, and Mark ($+$) means the predicted data. \label{asa_pre_2}}
\end{figure}

In addition, we define a maximum sampling period ($T_{s}^{max}$) to prevent the problem to miss disaster alarms due to the excessively extended sampling period. All sampling periods (${}^{\forall} T_{s}^{n}$) of proposed algorithms are smaller than $T_{s}^{max}$. 
\begin{IEEEeqnarray}{c}
{}^{\forall} T_{s}^{n}<T_{s}^{max} = \frac{\mathbf{G}^{margin}}{V_{c}^{event}} 
\label{T_a}
\end{IEEEeqnarray}
where $\mathbf{G}^{margin}$ is a gap ($D_n^{event}-D_n$) between event alarming threshold and current data. $V_{c}^{event}$ means an environment change velocity caused by disasters. Assuming a disaster occurs at a last sampling point, the maximum sampling period ($T_{s}^{max}$) is determined based on the time it takes for environment data to raise to the warning phase. Algorithm $1$ shows the pseudo code of ASA-m. 

\subsubsection{{Mean Velocity based Prediction (MVP)}}
This section introduces a novel adaptive filter, called MVP. ASA-m calculates optimal sampling periods by using MVP. It traces a measured behavior of change and predicts data. Figure~\ref{F16} and Table~\ref{notation_ASA} show operations and notations for MVP. Sensor nodes calculate a mean velocity ($V_{c}^{n}$, [Value/Sec]), i.e., the property of environmental change and monitored sources. To calculate the expected time for meaningful change ($B^{'}$), we define three types of mean velocities, $V_c^n$, $V_{c}^{pre}$, and $V_{c}^{post}$ as in (\ref{V_c_f})-(\ref{V_c}).
\begin{IEEEeqnarray}{c}\label{V_c_f}
V_{c}^{n}= \frac{|D_{n}-D_{n-1}|}{T_{s}^{n}}=(|D_{n}-D_{n-1}|)f_{s}^{n} \\
V_{c}^{pre}=\frac{V_{c}^{n-1}+V_{c}^{n}}{2},\quad V_{c}^{post}=\frac{V_{c}^{n}+V_{c}^{n+1}}{2}
\label{V_c}
\end{IEEEeqnarray}
{where $D_n$ is $n$-th sampled data. $n$ is a step counter.} $V_{c}^{pre}$ refers to the mean velocity between a latest target and a latest sampled data. $V_{c}^{post}$ represents the mean velocity between a current target and a current sampled data. Equation~(\ref{V_n+1}) shows the next mean velocity ($V_{c}^{n+1}$) applying the variability ($\Delta V$) in the monitoring environment as below:
\begin{IEEEeqnarray}{c}
V_{c}^{n+1}= V_{c}^{n}+\Delta V=V_{c}^{n}+(V_{c}^{n}-V_{c}^{n-1})
\label{V_n+1}
\end{IEEEeqnarray}

The next mean velocity follows a prior trend to trace the change of environment. Using these parameters, ASA-m calculates the next sampling period in the direction of reducing error. ASA-m separates the trend of environmental change into two cases (i.e., increasing and decreasing) as follows: 
\begin{IEEEeqnarray}{c}
T_{s}^{n}=\left\vert\frac{D_{n}-D_{n-1}}{V_{c}^{n}}\right\vert
\end{IEEEeqnarray}
\textbf{Case 1 :} The increasing trend ($D_{n}>D_{n-1}$)
\begin{IEEEeqnarray}{c} \label{T_post_I}
T_{s}^{post}=\frac{2(B^{'}-r_{n})}{|3V_{c}^{n}-V_{c}^{n-1}|},\quad T_{s}^{pre}=\frac{2r_{n-1}}{|V_{c}^{n}+V_{c}^{n-1}|} 
\end{IEEEeqnarray}
\textbf{Case 2 :} The decreasing trend ($D_{n}<D_{n-1}$).
\begin{IEEEeqnarray}{c} \label{T_post_D}
T_{s}^{post}=\frac{r_{n}}{\vert 3V_{c}^{n}-V_{c}^{n-1}\vert},\quad T_{s}^{pre}=\frac{2(B^{'}-r_{n-1})}{\vert V_{c}^{n}+V_{c}^{n-1}\vert}
\end{IEEEeqnarray}
{where $r_n$ is the distance between sampled ($D_n$) and below target ($B_s$) data. $B^{'}$ is a meaningful change.}

ASA-m defines the sensitivity tuner ($\alpha$) in (\ref{D_B}). It is a parameter to control the allowable error of sampling operation. If a sampling error is over $\alpha B^{'}$, ASA-m recalculates the next sampling rate. Sensor nodes predict the next sampling period while eliminating the prior error. 
\begin{IEEEeqnarray}{c}
\vert D_{n}-B_{n}\vert \geq \alpha B^{'}
\label{D_B}
\end{IEEEeqnarray}
where $B_n$ is $n$-th target data. $\alpha$ is a parameter within a range of ($0-0.5$) by system requirements. When $\alpha$ is small, it increases both computational complexity and estimation accuracy of sampling periods.

The sensor node computes the next sampling period ($T_{s}^{ASA-m}$) using the meaningful change ($B^{'}$) and estimated mean velocity ($V_c^{n+1}$). ASA-m differently calculates the next sampling period ($T_{s}^{ASA-m}$), according to three sampling conditions: Stable ($\vert D_{n}-B_{n}\vert < \alpha B^{'} $), Too Short ($\vert B_{n}-B_{n-1}\vert=0$), and Too Long ($\vert B_{n}-B_{n-1}\vert > B^{'}$). The formulas of increasing trend case are listed in (\ref{T_n+1})-(\ref{SF_next}). In decreasing trend case, the equation part ($2B^{'}-r_{n}$) in (\ref{T_n+1})-(\ref{SF_next}) changes to ($B^{'}+r_{n}$). Other parts are the same. 
\begin{IEEEeqnarray}{c}
T_{s}^{ASA-m}=
\begin{cases}
\frac{B^{'}}{|B_{n}-B_{n-1}|}T_{s}^{n}, & \mbox{if }\vert B_{n}-B_{n-1}\vert > B^{'}\\
\frac{B^{'}}{\vert V_{c}^{n+1}\vert}, & \mbox{if }\vert D_{n}-B_{n}\vert < \alpha B^{'} \\
\frac{2B^{'}-r_{n}}{\vert V_{c}^{n+1}\vert}, & \mbox{otherwise}
\end{cases}
\label{T_n+1}
\end{IEEEeqnarray}

Based on the Equation~(\ref{T_n+1}), The sampling rate of ASA-m is defined as given below: 
\begin{IEEEeqnarray}{c}
f_{s}^{ASA-m}=
\begin{cases}
\frac{\vert B_{n}-B_{n-1}\vert}{B^{'}}f_{s}^{n}, & \mbox{if }\vert B_{n}-B_{n-1}\vert > B^{'} \\
\frac{2\vert D_{n}-D_{n-1}\vert}{B^{'}}f_{s}^{n} &\mbox{if }\vert D_{n}-B_{n}\vert < \alpha B^{'}\\
\quad-\frac{\vert D_{n-1}-D_{n-2}\vert}{B^{'}}f_{s}^{n-1}, \\
\frac{2\vert D_{n}-D_{n-1}\vert}{2B^{'}-r_{n}}f_{s}^{n} & \mbox{otherwise}\\
\quad-\frac{\vert D_{n-1}-D_{n-2}\vert}{2B^{'}-r_{n}}f_{s}^{n-1}, 
\end{cases}
\label{SF_next}
\end{IEEEeqnarray}

Following equations~(\ref{T_n+1})-(\ref{SF_next}), ASA-m adjusts the next sampling rate. In Too long case, ASA-m misses some target data. The next sampling rate increases as much as the existing ASA-m sampling rate times the number of missed target data. Otherwise, in Too Short case, a sensor node collects the same data at the next sampling period. ASA-m recalculates the sampling rate by using the measured mean velocity. 

Figure~\ref{F17} shows a brief result of temperature sampling using ASA-m, and it shows only sampled data by a sensor node. In contrast, Figure~\ref{asa_pre_2} shows the data plane on a server-side that includes both sampled and estimated data. Mark ($\bigcirc$) means the sampled data, and mark ($+$) denotes the predicted data. Figure~\ref{F17} shows that sensor nodes accurately collect target data through ASA-m sampling rate because MVP includes the transition trend of disaster sources. ASA-m makes all the sampled data of sensor nodes worthwhile because the server compensates sampling errors occurred by sensor nodes.

\subsection{Continuous Adaptive Sampling Algorithm (CASA)}
The proposed protocol adopts CASA to the sensor nodes that still suffer from energy deficiency despite of ASA-m. Sensor nodes save more energy by adopting CASA. CASA is a simple method to use predicted data when the variability of data is very small ($|\Delta V| < \mathbf{M}$). {CASA factor ($\mathbf{M}$) is an upper bound of allowable variability for judging linear change.} In this condition, a sensor node accurately predicts the next sampling rate by using a linear trend estimation. Thus, the sensor node skips one step. The server in a network generates the predicted data. Using this advantage, sensor node saves much energy. The advantage makes sensor nodes to endure their power constraints. Sensor nodes recharge a small amount of energy if their harvesting quality is high.

The equations~(\ref{casa_f})-(\ref{casa}) show how to calculate the sampling rate of CASA and CASA factor ($\mathbf{M}$). {CASA sets the sampling period to plus the ASA-m sampling period and the next ASA-m sampling period in (\ref{casa_f}) when it satisfies the CASA condition ($\vert \Delta V\vert< \mathbf{M}$).} Equation~(\ref{casa_1}) shows that the mean velocity after next is within a range of allowable error ($\epsilon$). CASA defines the CASA factor ($\mathbf{M}$) from (\ref{casa_1}) as in (\ref{casa}). $\mathbf{M}$ means the bound to satisfy condition (\ref{casa_1}). That is a criterion for the judgment of linear change. Sensor nodes independently set a value of $\mathbf{M}$.
\begin{IEEEeqnarray}{c}\label{casa_f}
 f_{s}^{CASA}=\frac{1}{T_{s}^{n+1}+T_{s}^{n+2}},\quad \mbox{if }\vert \Delta V\vert< \mathbf{M} \\
 (1-\epsilon)V_{c}^{n} < V_{c}^{n+2} < (1+\epsilon)V_{c}^{n} \label{casa_1}\\
 \mathbf{M}= \left\vert\frac{\epsilon V_{c}^{n}}{2}\right\vert 
\label{casa}
\end{IEEEeqnarray}
{where $T_s^{n+1}$ is ASA-m sampling period. $T_s^{n+2}$ denotes the next ASA-m sampling period. $\epsilon$ means the allowable error of estimated mean velocity ($V_c^{n+2}=V_c^n+2\Delta V$) from (\ref{V_n+1}).}
\begin{figure}[!t]
\setlength{\fboxsep}{0pt}%
\setlength{\fboxrule}{0pt}%
\centerline{\includegraphics[width=8.8cm]{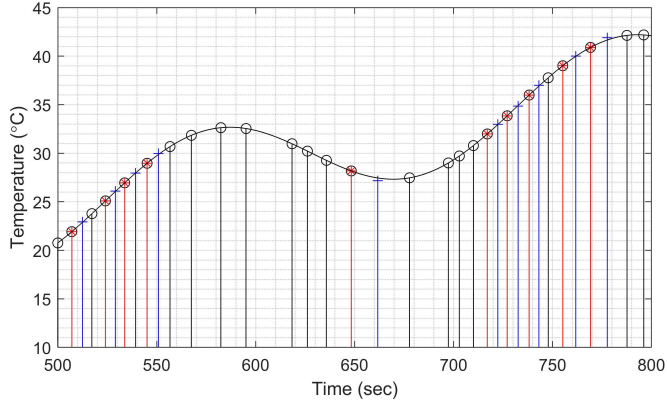}}
\caption{The sampling result using ASA-m with CASA based on a dynamic transition data set, when $\alpha$ is 0.2 and $B^{'}$ is 1 Celsius ($\bigcirc$ = sampled data, $\circledast$ = operation point of CASA and send the class information to the server, $+$ = predicted data). \label{F19}}
\end{figure}
Sensor nodes using CASA send their class information with data to the server. When the server receives the information of class 1, the server generates the predicted data after the predicting rate ($f_{s}^{pd}$) in (\ref{casa_pre}).
\begin{IEEEeqnarray}{c}
f_{s}^{pd}=\frac{1}{T_{s}^{n+1}}
\label{casa_pre}
\end{IEEEeqnarray}
{where $f_{s}^{pd}$ is the predicting rate calculated on the server-side.} It is same with ASA-m sampling rate. CASA adds a condition for reliable operation. Equation~(\ref{casa_end}) introduces the condition for eliminating algorithmic error of CASA in Algorithm $2$.
\begin{IEEEeqnarray}{c}
|V_{c}^{n}-V_{c}^{n-1}| \neq 0
\label{casa_end}
\end{IEEEeqnarray}

\begin{table}[!t]
\label{table_casa}
\centering
% Some packages, such as MDW tools, offer better commands for making tables
% than the plain LaTeX2e tabular which is used here.
\begin{tabular}{m{8.3cm}}
\hline
$\mathbf{Algorithm\,2:}$ Continuous\,adaptive\,sampling\,algorithm\,(CASA)\\
\hline
$\mathbf{Input \,\;\,\,:}$ Adaptive sampling rate ($f_{s}^{ASA-m}$) \\
$\mathbf{Output:}$ Continuous adaptive sampling rate ($f_{s}^{CASA}$) and estimated data\\[3pt]
$\mathbf{Sensor\,node\,process:}$ \\
$\mathbf{if}$ The variability ($\Delta V$) is very small $(|V_{c}^{n}-V_{c}^{n-1}|<\mathbf{M}\, \&\&\, |V_{c}^{n}-V_{c}^{n-1}| \neq 0)$ \\
\quad $\mathbf{Calculate}$ the next ASA-m period ($T_s^{n+2}$) \\
\quad$V_{c}^{n+2}=V_{c}^{n+1}+(V_{c}^{n}-V_{c}^{n-1})$ ; (in~(\ref{V_n+1}))\\
\quad$T_{s}^{n+2}=B^{'}/|V_{c}^{n+2}|$; (in~(\ref{T_n+1}))\\
\quad $\mathbf{Plus}$ a ASA-m period ($T_s^{n+1}$), the next ASA-m period ($T_s^{n+2}$) \\
\quad$T_{s}^{CASA}=T_{s}^{n+1}+T_{s}^{n+2}$; (in~(\ref{casa_f}))\\
\quad$f_{s}^{CASA}=1/T_{s}^{CASA}$;\\
\quad$V_{c}^{n+1}=V_{c}^{n+2}$;\\
\quad$B_{n}=round(B_{n}+(V_{c}^{n+1}*T_{s}^{n+1})$);\\[3pt]
$\mathbf{Server\,process:}$ Generate the $1$ predicted data after $f_{s}^{pd}$ as in~(\ref{casa_pre})\\
$\mathbf{if}$ Receiving class information 1\\ 
\quad $D_{pre}=\left(D_{n}+(V_{c}^{n+1}*T_{s}^{n+1})\right)$; \\
\hline
\end{tabular}
\end{table}

Figure~\ref{F19} shows an brief result of the performance of CASA. Mark($\circledast$) denotes the sampled data from a sensor node, and mark ($+$) means the predicted data from a network server. Algorithm $2$ shows the pseudo code. CASA contributes to improving the self-sustainability of sensor nodes and the accuracy of data prediction.

\subsection{Recoverable Adaptive Sampling Algorithm (RASA)}
\begin{figure}[!t]
\setlength{\fboxsep}{0pt}
\setlength{\fboxrule}{0pt}
\centerline{\includegraphics[width=8.8cm]{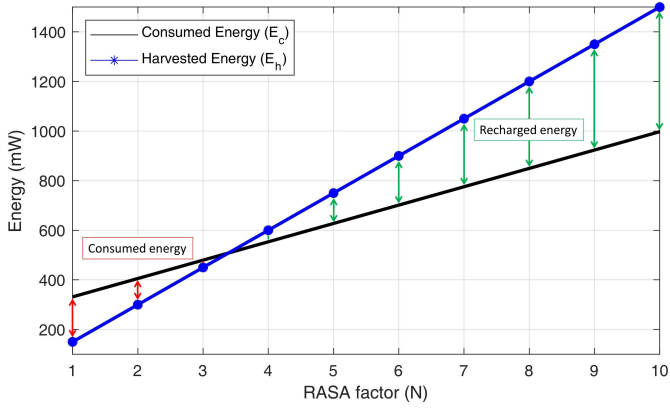}}
\caption{The comparison of harvested and consumed energy. {As selecting a RASA factor ($\textbf{N}$), it controls the amount of recharged energy.}\label{C_E}}
\end{figure}

RASA acts as an emergency energy recovery algorithm in ASMP. RASA recovers the energy state of a sensor node when it is in a crisis of energy exhaustion. Sensor nodes in ASMP need to manage their energy state autonomously since dead sensor nodes affect the whole network performance. The main concept of RASA is to make consumed energy ($E_{c}(i)$) smaller than EWMA of harvested energy ($\mathbf{E}[E_{h}^{i}]$). The formula shown in (\ref{rasa_1}) explains this condition. By adding formulas~(\ref{rasa_1})-(\ref{rasa}) to ASA-m, sensor nodes reduce sampling and transmitting operation by using the predicted data and RASA factor ($\mathbf{N}$). {RASA factor ($\mathbf{N}$) is a multiplier for extending a ASA-m sampling period of energy-hungry sensor nodes. RASA defines a sustainable time ($T_{sus}$) of sensor nodes in Equation~(\ref{eq:5}).} It shows the relation between sustainable time and RASA sampling rate based on (\ref{eq:4_f}). RASA extends $T_{sus}$.
\begin{IEEEeqnarray}{c} 
T_{sus}\propto \textbf{N}*\frac{E_{r}+(Q_{h}*P_{h}^{max})}{f_{s}^{ASA-m}*E_{sp}}
\label{eq:5}
\end{IEEEeqnarray}
{where energy definitions ($E_r$, $E_{sp}$, $Q_h$, and $P_h$) refer to Table~\ref{notation} in Section $\mathbf{V}$.} To increase $T_{sus}$ of sensor nodes, RASA manages such that sensor nodes satisfy the condition of (\ref{rasa_1}). Equation~(\ref{ewma}) represents EWMA of harvested energy. Using Equations from~(\ref{rasa_1})-(\ref{ewma}), we derive Equation~(\ref{rasa}).
\begin{IEEEeqnarray}{c} \label{rasa_1}
\mathbf{E}[E_{h}^{i}] > E_{c}(i)\\ 
\label{ewma}
\mathbf{E}[P_{h}^{i}] = \lambda*P_{h}^{i}+(1-\lambda)*\mathbf{E}[P_{h}^{i-1}]\\
\frac{\textbf{N}*\mathbf{E}[P_{h}^{i}]}{f_{s}^{ASA-m}} > E_{sl}+E_{w}+E_{ss}+E_{p}+E_{tx}
\label{rasa}
\end{IEEEeqnarray}
where $\lambda$ is a weighting factor in EWMA of harvested energy. {$\mathbf{E}[\bullet]$ is average value. Energy notations refer in Table~\ref{notation}.} To calculate the RASA factor ($\textbf{N}$), we define the energy consumption over time of a sensor node. Equation~(\ref{rasa_time_f})-(\ref{rasa_time}) show the Wake up ($T_w$) and Sleep ($T_{sl}$) time of a sensor node, where ASMP uses a sleep and wake up protocol.
\begin{IEEEeqnarray}{c}\label{rasa_time_f}
T_{w} = T_{ss}+T_{p}+T_{tx}\\
T_{sl} = (\textbf{N}*T_{s}^{ASA-m})-T_{w}
\label{rasa_time}
\end{IEEEeqnarray}
{where $T_s^{ASA-m}$ is ASA-m sampling period. $T_w$ is wake up time. $T_{sl}$ is sleep time. $T_{ss}$, $T_{p}$, and $T_{tx}$ are sensing, processing, and transmission time.}
Equations~(\ref{rasa_f_f})-(\ref{rasa_f}) show the calculation method of RASA factor ($\textbf{N}$). This is derived from the intersecting point of energy shown in Figure~\ref{C_E}. 
\begin{IEEEeqnarray}{c}\label{rasa_f_f}
\textbf{N} > \frac{(P_w-P_{sl})T_w+P_{ss}T_{ss}+P_p(l)T_p+P_{tx}T_{tx}}{(\mathbf{E}[P_{h}^{i}]-P_{sl})T_s^{ASA-m}}\\
f_{s}^{RASA} =\frac{f_{s}^{ASA-m}}{\textbf{N}}
\label{rasa_f}
\end{IEEEeqnarray}
{where power and time notations refer to Table~\ref{notation}.}
RASA starts when it is below a certain energy state ($\textbf{L}$). RASA factor ($\textbf{N}$) adjusts the energy recovering ratio because $\textbf{N}$ extends the sleep duration of a sensor node. Figure~\ref{C_E} shows the relation between harvested ($E_h$) and consumed ($E_c$) energy following the value of $\textbf{N}$. $\textbf{N}=\{1,2,3,...\}$ is an integer. RASA sampling rate should be smaller than the maximum frequency ($f_{max}^{RASA}$) to satisfy (\ref{rasa_1}). In ASMP, RASA transmits the value of $\textbf{N}$ to a server of the system. The server generates the $(\textbf{N}-1)$ numbers of predicted data for guaranteeing QoS as in server process of Algorithm $3$. Algorithm $3$ introduces the pseudo code of RASA. To minimize the number of predicted data, a sensor node sets the minimized value of $\textbf{N}$ like (\ref{rasa_pre}).
\begin{figure}[!t]
\setlength{\fboxsep}{0pt}
\setlength{\fboxrule}{0pt}
\centerline{\includegraphics[width=8.8cm]{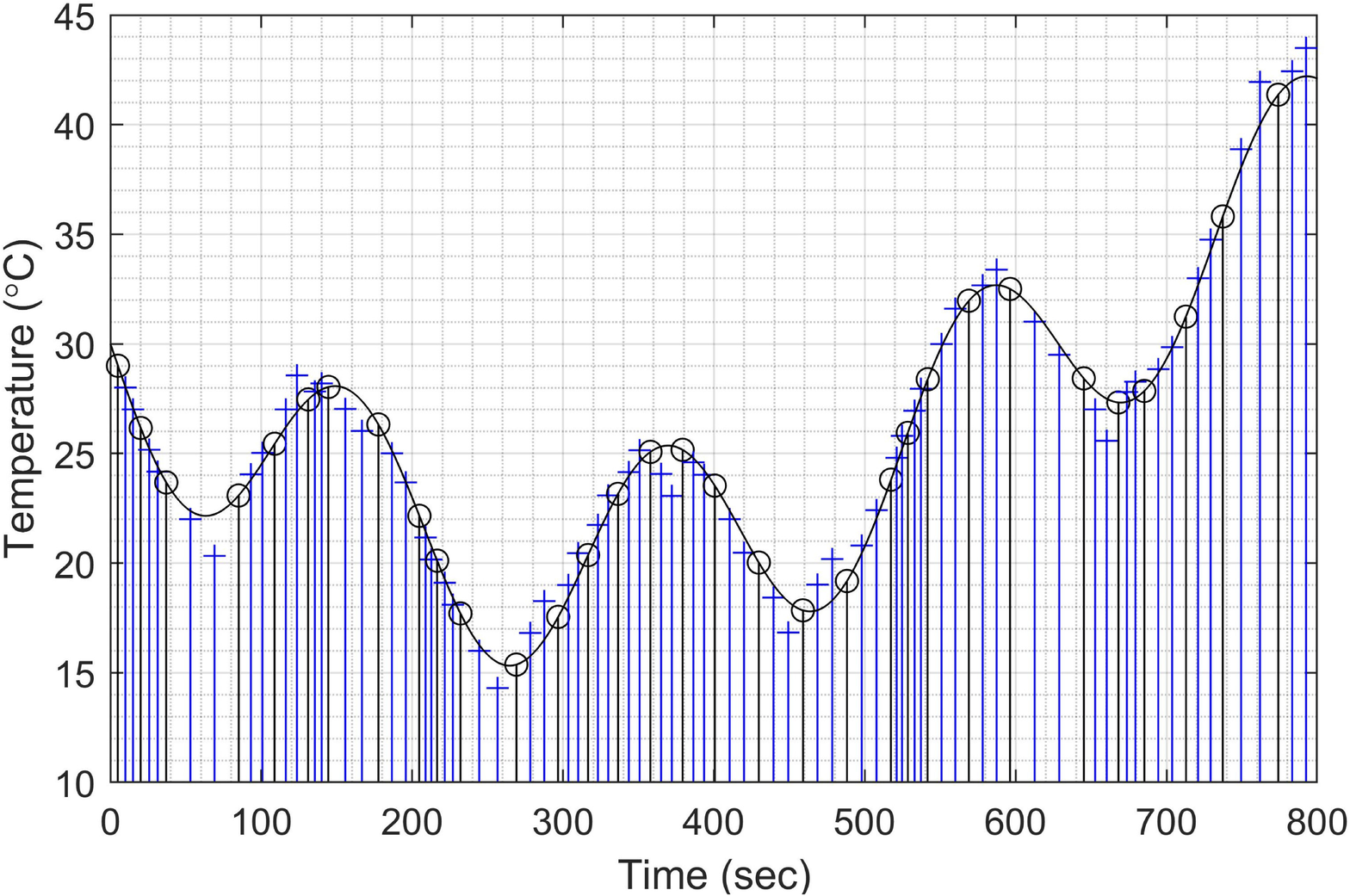}}
\caption{The sampling result using the ASA-m with RASA based on a dynamic transition data set, when $\alpha$ is 0.2 and $B^{'}$ is 1 Celsius, $\bigcirc$ = sampled data, and $+$ = predicted data.\label{F25}}
\end{figure}
\begin{IEEEeqnarray}{c}
f_{max}^{RASA} =\frac{f_{s}^{ASA-m}}{min(\textbf{N})}
\label{rasa_pre}
\end{IEEEeqnarray}

Figure~\ref{F25} shows an example of the operation of RASA. The sampled data is marked ($\bigcirc$), whereas the predicted data is marked ($+$). For disaster monitoring, the sampling rate of sensor nodes does not set lengthy periods so as not to miss the disaster event situation. We defines the event velocity ($V_{c}^{event}$). We also define the margin of the event ($\mathbf{G}^{margin}$) to solve the problem of miss-detection probability. The margin of the event refers to the difference between the alarm and sampled values. Using these definitions, the maximum sampling period is defined as Equation (\ref{event-margin}). RASA calculates the maximum sampling period ($T_s^{max}$) by dividing ($\mathbf{G}^{margin}$) into the event velocity ($V_{c}^{event}$). It is an empirical value when a disaster occurs.
\begin{IEEEeqnarray}{c}
T_{s}^{RASA} < T_{s}^{max} = \frac{\mathbf{G}^{margin}}{V_{c}^{event}} 
\label{event-margin}
\end{IEEEeqnarray}
where $V_{c}^{event}$ refers to the data-changing velocity when a disaster event occurs. RASA defines the sustainable time of sensor nodes using EWMA of harvested energy, as shown in (\ref{eq:5}). Considering the sustainable time of sensor nodes, RASA extends the lifetime of energy-deprived sensor nodes and provides a chance to recover and recharge.

\begin{table}[!t]
\label{table_rasa}
\centering
% Some packages, such as MDW tools, offer better commands for making tables
% than the plain LaTeX2e tabular which used here.
\begin{tabular}{m{8.3cm}}
\hline
$\mathbf{Algorithm\,3:}$ Recoverable\,adaptive\,sampling\,algorithm\,(RASA)\\
\hline
$\mathbf{Input:}$ Adaptive sampling rate ($f_{s}^{ASA-m}$) , consumed energy ($E_{c}(i)$) and EWMA of harvested energy ($\mathbf{E}[P_{h}^{i}]$) in (\ref{ewma})\\
$\mathbf{Output:}$ Recoverable adaptive sampling rate ($f_{s}^{RASA}$) and estimated data \\
\\[3pt]
$\mathbf{Sensor\,node\,process:}$ \\
$\mathbf{if}$ battery level is under $L$ ($E_{r}(i) < L$) \\
\quad $\mathbf{for}$ ($E_{c}(i) > E_{h}(i)\,\&\&\,\mathbf{G} < \mathbf{G}^{margin}$) \{ \\
\quad\quad $\mathbf{Calculate}$ a wake up ($T_{w}$) and sleep time ($T_{sl}$) of an Edge\\
\quad\quad$T_{w}=T_{p}+T_{ss}+T_{tx}$; (in~(\ref{rasa_time_f}))\\
\quad\quad$T_{sl}=N*T_{s}^{ASA-m}-T_{w}$; (in~(\ref{rasa_time}))\\
\quad\quad $\mathbf{Calculate}$ consumed ($E_{c}(i)$) and harvested ($E_{h}(i)$) energy \\
\quad\quad$E_{c}(i)=E_{sl}+E_{w}+E_{p}+E_{ss}+E_{tx}$ ; (in~(\ref{eq:2}))\\
\quad\quad$\mathbf{E}[P_{h}^{i}]=\lambda*P_{h}^{i}+(1-\lambda)*\mathbf{E}[P_{h}^{i-1}]$; (in~(\ref{ewma}))\\
\quad\quad$E_{h}(i)=\mathbf{E}[P_{h}^{i}]*N*T_{s}^{n}$; (in~(\ref{E_h}))\\
\quad\quad$\mathbf{Limit}$ a sampling period under the maximum margin in~(\ref{event-margin})\\
\quad\quad$\mathbf{G}=D_{n}+N*V_{c}^{event}*T_{s}^{n}$;\\
\quad\quad $\mathbf{Find}$ a value of $N$\\
\quad\quad $N=N+1$;\\
\quad$\mathbf{end}$ \\
\quad$\mathbf{Extend}$ the next sampling period\\
\quad$T_{s}^{RASA}=N*T_{s}^{ASA-m}$ ;\\
\quad$f_{s}^{RASA}=1/T_{s}^{RASA}$ ; (in~(\ref{rasa_f}))\\
\quad$B_{n}=round(B_{n}+(V_{c}^{n+1}*(N-1)*T_{s}^{ASA-m})$);\\
\\[3pt]
$\mathbf{Server\,process:}$ Generate $(N-1)^{th}$ predicted data every $T_{s}^{ASA-m}$ \\
$\mathbf{if}$ (Receiving class information $\textbf{N}$) \\
\quad$\mathbf{for}$ $k=[1:N-1]$\\
\quad\quad $D_{pre}^{k}=\left( D_{n}+(V_{c}^{n+1}*k*T_{s}^{ASA-m})\right)$; \\
\quad$\mathbf{end}$\\
\hline
\end{tabular}
\end{table}

\section{The mathematical energy model for energy harvesting WSNs}
To analyze the energy status of a sensor node, EH-WSNs require mathematical energy consumption and harvest models. In a real environment, sensor nodes measure the consumed and harvested energy by using current and voltage sensors when gathering environmental information. We need mathematical models to evaluate the performance of our protocol and to test the self-sustainability of sensor nodes. We calculate the remained, consumed, and harvested energies of sensor nodes. Zhou \textit{et al.}\cite{Zhou2011} introduces the energy models for each module such as processor, sensor, and transmitter. Srbinovski \textit{et al.}\cite{Srbinovski2016} introduces a comprehensive energy model for sensor nodes of WSNs. We propose an advanced energy consumption model including new parameters such as cluster size ($C$), processing load ($l$), and operation modes. 

We propose two important definitions to evaluate the energy state of a sensor node. First, we define the relation of a sampling rate ($f_{s}$) and energy consumption ($E_c(t)$). Second, each sensor nodes calculate the sustainable time ($T_{sus}$) by using this relation. The definitions are listed in (\ref{eq_f_e})-(\ref{eq:4}). \\
\textbf{Relation of a $E_{c}(t)$ and $f_{s}$ :}
\begin{IEEEeqnarray}{rCl} \label{eq_f_e}
 E_{Edge}(t)&\approx& E_{sp}f_{s} \\ 
 E_{Relay}(t)&\approx& 2E_{sp}f_{s} \\
 E_{Sync}(t)&\approx& CE_{sp}f_{s} 
\label{eq}
\end{IEEEeqnarray} 
\textbf{Sustainable time ($T_{sus}$) of a sensor node :}
\begin{IEEEeqnarray}{rCl} \label{eq:4_f} 
 \frac{E_{r}}{f_{s}E_{sp}}&=&T_{sus}\quad < Edge\,mode> \\
 \frac{E_{r}}{2f_{s}E_{sp}}&=&T_{sus}\quad <Relay\,mode> \label{eq:4}\\
 \frac{E_{r}}{Cf_{s}E_{sp}}&=&T_{sus}\quad <Sync\,mode> \label{eq_32}
\end{IEEEeqnarray}
{where $E_{sp}$ denotes the energy consumed by once sampling process in (\ref{eq:3_f}). The energy model by mode is in (\ref{E_sleep})-(\ref{E_sync}).} $C$ refers to the cluster size of EH-WSNs. The cluster size is closely related to the communication energy consumption of relay and sync modes because all sensor nodes in the cluster transmit their information to the sync (cluster header).

\begin{table}[t]
\renewcommand{\arraystretch}{1.3}
\caption{List of symbols}
\label{notation}
\centering
% Some packages, such as MDW tools, offer better commands for making tables
% than the plain LaTeX2e tabular which is used here.
\begin{tabular}{m{1.5cm}m{6.5cm}}
\hline
$\boldsymbol{Symbol}$ & $\boldsymbol{Description}$\\
\hline
$Q_{h}$&\text{Energy harvesting quality}\tabularnewline
$P_{h}^{i}$&\text{$i$-th harvesting power}\tabularnewline
$P_{h}^{max}$&\text{Maximum harvesting power}\tabularnewline
$E_{r}$&\text{Remained energy of a sensor node}\tabularnewline
$E_{sl}$&The sum of energy consumption caused by only sleep operations (Same with $P_{sl}$(Power), $T_{sl}$(Time))\tabularnewline
$E_{w}$&The sum of energy consumption caused by only wake-up operations (Same with $P_{w}$(Power), $T_{w}$(Time))\tabularnewline
$E_{p}$&The sum of energy consumption caused by only processing operations (Same with $P_{p}$(Power), $T_{p}$(Time))\tabularnewline
$E_{ss}$&The sum of energy consumption caused by only sampling operations (Same with $P_{ss}$(Power), $T_{ss}$(Time))\tabularnewline
$E_{rx}$&The sum of energy consumption caused by only receiving operations (Same with $P_{rx}$(Power), $T_{rx}$(Time))\tabularnewline
$E_{tx}$&The sum of energy consumption caused by only transmitting operations (Same with $P_{tx}$(Power), $T_{tx}$(Time))\tabularnewline
$E_{sp}$&The sum of energy consumed by once sampling process\tabularnewline
\hline
\end{tabular}
\end{table}
\begin{figure}[t]
\setlength{\fboxsep}{0pt}%
\setlength{\fboxrule}{0pt}%
\centerline{\includegraphics[scale=0.41]{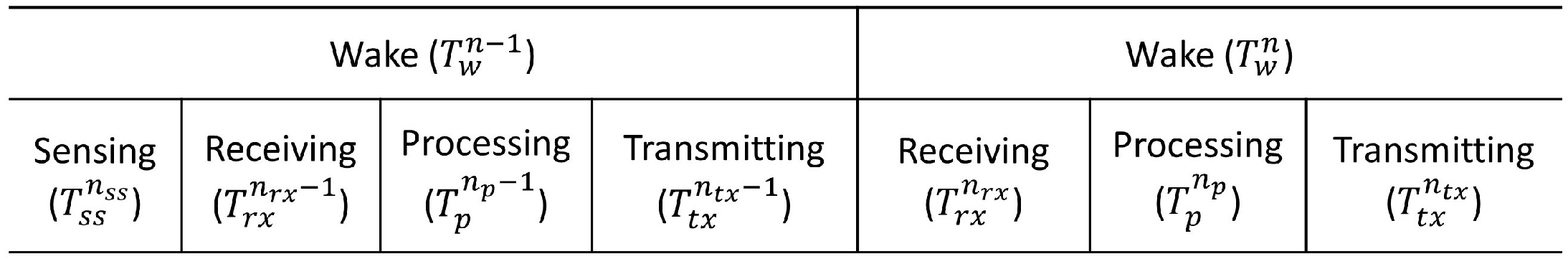}}
\caption{The time modeling example based on sampling periods when a sensor node operates as a sync mode.\label{F18}}
\end{figure}

Sensor nodes is divided into four modes (i.e., Sleep, Edge, Relay, and Sync) according to the role in a disaster-monitoring system. Sleep mode is the state that does not have any role in the network. In ASMP, sensor nodes operate to sleep mode ($T_{sl}$) when they are in a sampling period except the wake-up time ($T_{sl}=T_{s}-T_{w}$) in (\ref{T}). Secondly, Edge mode is the state that the sensor node plays a role on the edge of EH-WSNs. Sensor nodes in the edge conduct two roles, sampling and transmitting data to sync nodes or relay nodes. Thirdly, in Relay mode, sensor nodes conduct an additional role to pass the data from other sensor nodes to the sync node. Furthermore, Relay sensor nodes conduct the intermediate role between edge nodes and sync nodes when the communication coverage of edge nodes is short. Finally, in Sync node, sensor nodes collect all the data around sensor nodes and transmit them to gateways or a system server. 

Moreover, we subdivide energy consumption into each operations, such as sleep ($sl$), wake up ($w$), processing ($p$), sensing ($ss$), receiving ($rx$), and transmitting ($tx$), for more accurate energy analysis. Equations~(\ref{E_sleep})-(\ref{E_sync}) show the energy consumption model by mode during a sampling period ($T_s$) as presented below: \\
\textbf{Mode 1. Sleep :} Sleep
\begin{IEEEeqnarray}{c}
E_{sleep}=E_{sl}
\label{E_sleep}
\end{IEEEeqnarray} 
\textbf{Mode 2. Edge :} Wake-up + Sensing(including energy information) + Processing + Transmitting + Sleep
\begin{IEEEeqnarray}{c}
E_{Edge} = E_{w}+E_{ss}+E_{p}+E_{tx}+E_{sl}
\label{E_edge}
\end{IEEEeqnarray}
\textbf{Mode 3. Relay :} Wake-up + Sensing + Receiving + Processing + Transmitting + Sleep or Receiving + Transmitting + Sleep
\begin{IEEEeqnarray}{rCl}
E_{Relay} &=& E_{w}+E_{ss}+E_{rx}+E_{p}+E_{tx}+E_{sl}\\
E_{Relay} &=& E_{rx}+E_{tx}+E_{sl}
\label{E_relay}
\end{IEEEeqnarray}
\textbf{Mode 4. Sync :} Wake-up + Sensing + Receiving + Processing + Transmitting + Sleep or Wake-up + Receiving + Processing + Transmitting + Sleep
\begin{IEEEeqnarray}{rCl}
E_{Sync} &=& E_{w}+E_{ss}+E_{rx}+E_{p}+E_{tx}+E_{sl}\\
E_{Sync} &=& E_{w}+E_{rx}+E_{p}+E_{tx}+E_{sl}
\label{E_sync}
\end{IEEEeqnarray}
where energy definitions are in Table~\ref{notation}.

$E_{sp}$ in (\ref{eq:3_f}) means a energy consumption accompanying once sampling process of a sensor node. 
\begin{IEEEeqnarray}{c} \label{eq:3_f}
 E_{sp}=E_{w}^{'}+E_{ss}^{'}+E_{p}^{'}+E_{tx}^{'}+E_{sl}^{'} \\\label{eq:3_a}
 E_{w}^{'}=P_{w}T_{w},\,E_{ss}^{'}=P_{ss}T_{ss},\,E_{p}^{'}=P_{p}(l)T_{p}\\
 E_{tx}^{'}=P_{tx}T_{tx},\,E_{sl}^{'}=P_{sl}T_{sl}
\label{eq:3}
\end{IEEEeqnarray}
{where Wake up ($E_{w}^{'}$), Sensing ($E_{ss}^{'}$), Processing ($E_{p}^{'}$), Transmitting ($E_{tx}^{'}$), and Sleep ($E_{sl}^{'}$), represent the energy consumed according to once operation.} $l$ represents the processing load of a micro controller unit (MCU). 

Equations~(\ref{E_def_a})-(\ref{E_def}) show the definitions of energy consumption following each operations. A sensor node has diverse operational modes and conducts several roles in the network during a lifetime. A sensor node has a different count number by each operations. The time duration of each operations follows by the system schedule or sampling period.
\begin{IEEEeqnarray}{c}\label{E_def_a}
 E_{sl}(n_{sl})=P_{sl}\sum_{k=1}^{n_{sl}} T_{sl}^{k},\quad\;
 E_{w}(n_{w})=P_{w}\sum_{k=1}^{n_{w}} T_{w}^{k} \\\label{E_cl}
 E_{p}(n_{p},l)=\sum_{k=1}^{n_{p}}(P_{p}^{k}(l)T_{p}^{k}),\;\label{l}
 E_{ss}(n_{ss})=P_{ss}\sum_{k=1}^{n_{ss}} T_{ss}^{k} \\
 E_{rx}(n_{rx})=P_{rx}\sum_{k=1}^{n_{rx}} T_{rx}^{k},\quad\;
 E_{tx}(n_{tx})=P_{tx}\sum_{k=1}^{n_{tx}} T_{tx}^{k}
\label{E_def}
\end{IEEEeqnarray}
where $E_{sl}(n_{sl})$ is the cumulative sum of sleep operations of a sensor node. $n_{sl}$ is the counted number of sleep operations. {$P_{sl}$ and $T_{sl}$ denotes a sleep power and time period.}

Figure~\ref{F18} shows the time modeling case when a sensor node operates as a sync mode. Equation~(\ref{T_f}) shows the total operation time ($T(n)$) of a sensor node. Equation~(\ref{T}) shows the sum of the $n$-th sleep and wake up time, i.e., a sampling period. This slotted time ($T_s^n$) is defined by the sampling rate ($f_s^n$) in (\ref{SF_next}). The series of each operation composes the operating time of a sensor node like Figure~\ref{F18}.
\begin{IEEEeqnarray}{c}\label{T_f}
T(n) = \sum_{k=1}^{n_{sl}} T_{sl}^{k}+\sum_{k=1}^{n_{w}} T_{w}^{k} \\
T_s^n=T_{sl}^{n_{sl}}+T_{w}^{n_{w}}
\label{T}
\end{IEEEeqnarray}
{where $T_s^n$ is $n$-th sampling period. $T_{sl}^{n_{sl}}$ is $n_{sl}$-th sleep time. $T_{w}^{n_{w}}$ is $n_{w}$-th wake up time.}

$E_{c}(n)$ in (\ref{eq:2}) is the cumulative sum of all operations performed for lifetime of a sensor node. $E_{c}(n)$ derives the sum of cumulative sum of each operations in (\ref{E_def_a})-(\ref{E_def}). 
\begin{IEEEeqnarray}{rCl} 
 E_{c}(n)&=&E_{sl}(n_{sl})+E_{w}(n_{w})+E_{p}(n_{p})\nonumber\\
 &&+\>E_{ss}(n_{ss})+E_{rx}(n_{rx})+E_{tx}(n_{tx})
\label{eq:2}
\end{IEEEeqnarray}

We define the harvesting energy ($E_{h}$) and quality ($Q_h$). The harvesting energy varies on environmental conditions. $E_{h}(n)$ in (\ref{E_h}) refers to the total quantities of harvested energy. To accurately calculate the harvesting energy, we define a harvesting power differently for each sampling period so that it includes the harvesting environment varied over time.
\begin{IEEEeqnarray}{c}
 E_{h}(n)=\sum_{i=1}^{n}P_{h}^{i}T_s^i
\label{E_h}
\end{IEEEeqnarray}
{where $P_{h}^{i}$ is a harvesting power. $P_{h}^{i}T_s^i$ represents the harvested energy quantities during the $i$-th time period.} We define harvesting quality ($Q_h$) in (\ref{eq:Q_f}) so that sensor nodes recognize their own harvesting performance autonomously. This entails the ratio of the $i$-th and maximum harvesting power ($P_h^{max}$). Equation~(\ref{eq:Q}) shows the total harvested energy quantities including the variable of harvesting quality.
\begin{IEEEeqnarray}{c}\label{eq:Q_f}
 Q_h(i)=\frac{P_{h}^{i}}{P_{h}^{max}} \\
 E_{h}(n)=\sum_{i=1}^{n}Q_{h}^{i}P_{h}^{max}T_s^i
\label{eq:Q}
\end{IEEEeqnarray}

Using the mathematical models of consumed ($E_c(n)$) in (\ref{eq:2}) and harvested ($E_h(n)$) energy in (\ref{eq:Q}), the equation for a remained energy ($E_{r}(n)$) derives in (\ref{eq:1}). The sensor node mathematically analyzes own remained energy by using (\ref{eq:1}). 
\begin{IEEEeqnarray}{c}\label{eq:1} 
 E_{r}(n) = E_{r}(n-1)-E_{c}(n) + E_{h}(n)\\
 E_{r}(n=0) = E_{initial}
\label{E_ini}
\end{IEEEeqnarray}
{where $E_{initial}$ is the initial energy state of a sensor node.}

\section{Design for a self-sustainable disaster-monitoring system}
In this section, we introduce how to organize a self-sustainable disaster monitoring system. We summarize the required performances of a network and sensor nodes. Sub-sections explain about our disaster monitoring system.

\begin{figure}[t]
\setlength{\fboxsep}{0pt}
\setlength{\fboxrule}{0pt}
\centerline{\includegraphics[width=8.8cm]{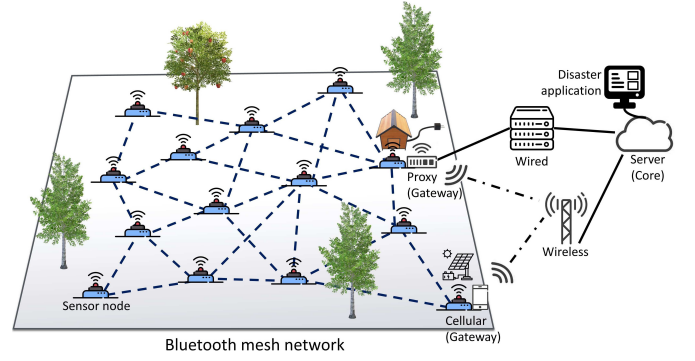}}
\caption{The simulation topology and system architecture using a internet connected Bluetooth mesh network.\label{simul_topo}}
\end{figure}

\subsection{The wireless network for a self-sustainable disaster-monitoring system}
In this section, we suggest a method to design a more stable and robust wireless network for disaster monitoring. It shows the conditions of our simulation. Sensor nodes are displayed in harsh places, such as mountains and underground, for gathering data related to disasters. Figure~\ref{simul_topo} shows a simulated system model, such as network topology and architecture. We use a Bluetooth mesh and Internet connected network because it has robustness, resilience, and self-organizability even at a low cost. End, Relay, and Sync nodes are in the Bluetooth network. Each node transmits gathered data to the server through gateways, which are connected wired or wireless. Gateways act as a network hub in the auxiliary power or power accessible location. This is the method to cover a wide forest and to monitor this area in real-time.

Firstly, sensor nodes sample and transmit the data at the same time for reliable disaster monitoring system. The fixed sampling method causes congestion and collision of data traffic, as well as the re-transmission of data because all sensor nodes transmit the data every same sampling periods ($T_s$). Otherwise, ASMP naturally disperses the network connection attempts of sensor nodes because they have all different sampling frequencies. This method also provides some advantages to reduce the congestion and collision of traffic, as well as energy consumption by re-transmission. These advantages improve QoS performances of the network.

Furthermore, the network uses a LEACH, which is one of cluster head selection algorithms for maintaining energy fairness, and a sleep and wake-up protocol for reducing wasted energy consumption. ASMP has the additional advantage to operate very effectively a sleep and wake-up protocol due to the characteristics of ASMP. Additionally, we use the concept of the self-organizing network. We organized the network by using the open source hardware for the compatibility, scalability, and interoperability of EH-WSNs.

\subsection{Sensor nodes for a self-sustainable disaster-monitoring system}
This section discusses how to design sensor nodes that minimize energy consumption for self-sustainability. Energy consumption involves three main parts, processor, communication, and sensor. We organized the system considering the energy consumption and functional properties of the open source hardware (OSHW) based on Table~\ref{table_2} and \ref{table_3}. We filled in the Table~\ref{table_2} and \ref{table_3} with reference to each datasheet written by manufacturers.

Firstly, monitoring hardware has to consider the processor because the processing energy is one of the main consumption parts. Table~\ref{table_2} shows the power consumption of MCUs, according to the operating status. We select the smallest energy consumption model (i.e., ATmega4809) for a simulation since low power consumption MCU is required.

\begin{table}[!t]
\renewcommand{\arraystretch}{1.3}
\renewcommand{\multirowsetup}{\centering}
\caption{Comparison the energy consumption and specifies of microprocessors}
\label{table_2}
\centering
% Some packages, such as MDW tools, offer better commands for making tables
% than the plain LaTeX2e tabular which is used here.
\begin{tabular}{ c c >{\centering}m{1.2cm}>{\centering}m{1.2cm}>{\centering}m{1.2cm}}
\hline
\multirow{2}{1.2cm}{Models} & \multirow{2}{1.5cm}{Performance\\(Max)} & \multicolumn{3}{c}{Power Consumption($mW$)}\tabularnewline
\cline{3-5}
&& \multicolumn{1}{c}{Sleep$^{1}$} & \multicolumn{1}{c}{Idle$^{2}$} & \multicolumn{1}{c}{MAX$^{3}$}\tabularnewline
\hline
ATmega128& $20MHz$ & $15\times 10^{3}$& $40$ & $200$ \tabularnewline
\hline
ATmega328p & $20MHz$ & $3\times10^{-4}$ & $6$ & $71.5$\tabularnewline
\hline
ATmega2560 & $16MHz$ &$3\times 10^{-3}$ & $13.5$ & $126.5$ \tabularnewline
\hline
ATmega1284P & $20MHz$ & $45\times 10^{-5}$ & $5.5$ & $104.5$ \tabularnewline
\hline
ATmega168P& $20MHz$ & $3\times 10^{-4}$ & $8.5$ & $93.5$ \tabularnewline
\hline
ATmega4809 & $20MHz$ & $3\times10^{-4}$ & $3.5$ & $52.5$ \tabularnewline
\hline
\end{tabular} \\[3pt]
\centering\footnotesize Please refer to the datasheets for various modules\\
\raggedleft\footnotesize $^{1}$ Based on 3V and WDT disabled\\
\raggedleft\footnotesize $^{2}$ Based on 5V and typical current\\
\raggedleft\footnotesize $^{3}$ Based on maximum voltage and performance\\
\end{table}
\begin{table}[!t]
\renewcommand{\arraystretch}{1.3}
\renewcommand{\multirowsetup}{\centering}
\caption{Energy consumption comparison of wireless communication technologies}
\label{table_3}
\centering
% Some packages, such as MDW tools, offer better commands for making tables
% than the plain LaTeX2e tabular which is used here.
\begin{tabular}{>{\centering}m{1.4cm}>{\centering}m{1.2cm}>{\centering}m{1.5cm}>{\centering}m{0.5cm}>{\centering}m{0.7cm}>{\centering}m{0.9cm}} 
\hline
\multirow{2}{1.4cm}{Technology (Model)} & \multirow{2}{1.2cm}{Data Rate ($kbps$)} & \multirow{2}{1.5cm}{Coverage ($m$)} & \multicolumn{3}{c}{Power Consumption($W$)}\tabularnewline
\cline{4-6}
& & & \multicolumn{1}{c}{Sleep} & \multicolumn{1}{c}{Rx} & \multicolumn{1}{c}{Tx}\tabularnewline
\hline
Zigbee\par\scriptsize{(XBee3 PRO)} & $250$ & $90$\scriptsize{(Indoor)} \par $3.2k$\scriptsize{(Outdoor)}& $7.2\mu$ & $61.2m$ & $486m$ \tabularnewline
\hline
BLE v5.0\par\scriptsize{(BM70)} & $2000$ & $50$ & $9\mu$ & $9.2m$ & $11.6m$\tabularnewline
\hline
LoRa\par\scriptsize{(SX1276)} & $300$ & $10k$ & $5.4\mu$ & $40.7m$ & $111m$\tabularnewline
\hline
\end{tabular} \\[3pt]
\centering\footnotesize {Please refer to the datasheets for various modules}\\
\end{table}

Secondly, a sensor node hardware for monitoring has considered the communication modules since the communication energy consumption contributes the biggest impact to the lifetime of EH-WSNs. Table~\ref{table_3} shows the energy consumption comparison of wireless communication technologies. The proposed sensor node uses Bluetooth communication, considering energy consumption and coverage. 

Finally, the sensing energy consumption of a sensor node is also one of the main energy consumption parts. Sensor modules are selected and used following the sensing resolution of QoS requirements and the ability to gather data.  

\begin{table}[t]
\renewcommand{\arraystretch}{1.3}
\caption{Design of a disaster monitoring sensor node}
\label{simulation_model}
\centering
\begin{tabular}{>\centering m{1.8cm}>\centering m{1.4cm}>\centering m{2cm}>\centering m{2cm}}
\hline
Modules & Sleep[$\mu W$] & Active[$mW$] & Time[$ms$] \tabularnewline
\hline
\multirow{2}{1.8cm}{Atmega4809 (nano envy)} & $0.3$ & $3.5$ & $T_{w}^{i}$\tabularnewline
\cline{2-4}
&\multicolumn{2}{c}{$P_{p}(l) = 0.5l+2.7^{a}$} & $T_{p}^{i}$ \tabularnewline
\hline
BM70 \par (BLE v5.0)] & $9$ & $9.2$\,(Rx), \par $11.6$\,(Tx) & $T_{rx}^{i}, T_{tx}^{i}=2.5^{b}$\tabularnewline
\hline
SC10050 (Solar) & - & $500*Q_{h}(n)$ & $T(n)$\tabularnewline
\hline 
DS18B20 (Temperature) & $5.5$ & $7.5$ & $T_{ss}\approx 1.4$ \tabularnewline
\hline 
KY-026 (Flame) &1&5& $T_{ss}\approx 1.4$\tabularnewline
\hline 
LSM303 (Acceleration) & $3.6$ & $0.36$ & $T_{ss}\approx 1.4$ \tabularnewline
\hline 
D7S (Vibration) &$450$&$1.5$& $T_{ss}\approx 1.4$ \tabularnewline
\hline 
K-0135(Water level sensor) &-&$100$& $T_{ss}\approx 1.4$\tabularnewline
\hline 
Battery pack (Li-Po) & - & $5000mAh$\,($7.4V$) & - \tabularnewline
\hline 
\end{tabular} \\[3pt]
\centering\footnotesize Reference from the datesheet of each model\\
\raggedleft \footnotesize{$^{a}$ Defined processing power according to the load from datasheet}\\
\raggedleft \footnotesize{$^{b}$ Based on maximum payload size of bluetooth v5.0}
\end{table}

\begin{table}[t]
\renewcommand{\arraystretch}{1.3}
\caption{Parameter list for a sensor node}
\label{parameter list}
\centering
% Some packages, such as MDW tools, offer better commands for making tables
% than the plain LaTeX2e tabular which is used here.
\begin{tabular}{m{3.5cm}>\centering m{1.2cm}>\centering m{1.3cm}>\centering m{1cm}}
\hline
Parameters & Symbol & Unit & Values\tabularnewline
\hline
\text{MCU Performance} & - & MHz &$20$\tabularnewline
\text{Communication coverage} & - & Meter & $50$\tabularnewline
\text{Data rate} & - & \text{Mb/s} &$2$\tabularnewline
\text{EWMA weight} & $\lambda$ & - &$0.001$\tabularnewline
\text{CASA allowable error} & $\epsilon$ & - & $0.3$\tabularnewline
\hline
Dynamic transition case \tabularnewline
\hline
\text{Meaningful change}& $B^{'}$ & \text{Celsius} & $1$\tabularnewline
\text{ASA sensitivity tuner} & $\alpha$ & - & $0.2$\tabularnewline
\text{Initial sampling rate} & $f_s$ & Hz & $0.2$\tabularnewline
\text{Maximum sampling period} & $T_{s}^{max}$ & sec & $\infty$\tabularnewline
\hline
Open access case \tabularnewline
\hline
\text{Meaningful change}& $B^{'}$ & \text{Celsius} & $0.2$\tabularnewline
\text{ASA sensitivity tuner} & $\alpha$ & - & $0.4$\tabularnewline
\text{Initial sampling rate} & $f_s$ & mHz & $4.76$\tabularnewline
\text{Maximum sampling period} & $T_{s}^{max}$ & min & $10$\tabularnewline
\hline
\end{tabular}
\end{table}

\subsection{Environments for the simulation}
We propose the method to organize sensor nodes for disaster monitoring, such as presented in Table~\ref{simulation_model}. It shows the energy consumption and operation time of each module. The values in Table~\ref{simulation_model} are collected from the datasheet or measured by experiments. Using these factors, we set up the simulation environment and evaluate the proposed protocol. The designed sensor nodes used the low power board and communication module, Arduino Nano envy and BM70 (Bluetooth v5.0), respectively. Sensor nodes use the battery pack combined of two small-sized batteries ($5000mAh, 3.7V$). In this simulation, we set the amount of harvested energy to $30mW$ based on the annual harvested energy statistics of Seoul, Republic of Korea. Day and nighttime are assumed to be equal to 12 h. Sensor nodes gather data related to wildfire, earthquake, flooding, and so on, using ASMP. We assume the sensing time to acquire the data of each sensor is the same because the sensing time of the sensor is related to the clock frequency of an MCU. The sensor node follows the fastest sampling rate when the system demands to monitor multiple environments. Table~\ref{parameter list} shows the parameter list for this simulation.

In this study, we conduct the simulations using two data sets such as dynamic transition and open access data set to show the contributions of our work. The open access data were achieved from the Korea Meteorological Administration (KMA)\footnote{https://data.kma.go.kr}. And, the dynamic transition data set were derived to demonstrate the performances of our contributions in a dynamic environment. The maximum and minimum value of dynamic data set are $43$ and $16$ Celsius. The dynamic data set has a large variance in temperature and changes frequently. Open access data is a empirical data set measured every 1 minute by KMA. The maximum and minimum value are $20.4$ and $4.4$ Celsius. It has smaller variance in temperature than a dynamic set, but the amount of data is sufficient.

\begin{figure}[t]
\setlength{\fboxsep}{0pt}
\setlength{\fboxrule}{0pt}
\centerline{\includegraphics[width=8.8cm]{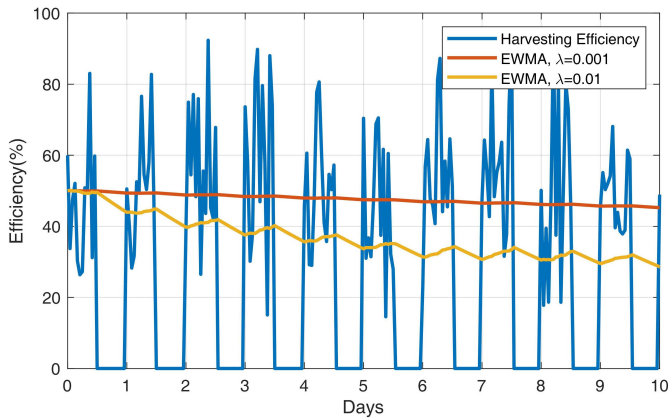}}
\caption{The mathematically modeled random solar energy harvesting quality based on a KMA open access data. It shows the EWMA of harvested energy according to $\lambda$.\label{F20}}
\end{figure}

\subsection{Random model of a solar energy harvest}
Harvesting energy is difficult to estimate since all harvesting techniques have high dependencies on environment change and their location. Therefore, we approaches energy harvesting based on a random model. Figure~\ref{F20} shows the random model of energy harvesting efficiency utilizing the solar radiation statistics of Seoul\cite{Lee2013a}. The mean and variance values derive from the meteorological statistics of Seoul provided by KMA. Following these data, we set that the average annual solar radiation quantity in Seoul is approximately $5kWh/m^2/day$. The energy harvesting techniques, such as solar and wind, have much bigger variations of harvested power than power line transfer. hence, the battery has to be large enough to control the variation of harvested power. Sensor nodes control their consumed energy using proposed adaptive algorithms according to the battery and harvesting state. They reliably operate using harvesting energy and battery as if a power line.

\begin{figure}[t]
\setlength{\fboxsep}{0pt}
\setlength{\fboxrule}{0pt}
\centerline{\includegraphics[width=8.8cm]{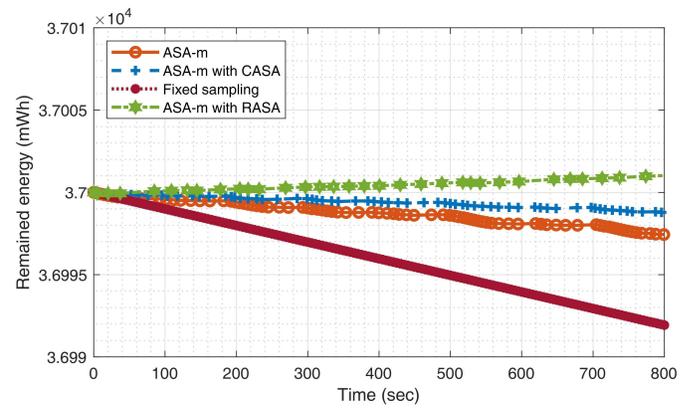}}
\caption{The energy status comparison of ASAs based on a dynamic transition data set.\label{ASA}}
\end{figure}

\begin{figure*}[!t]
\centering 
\begin{tabular}{cc}
\includegraphics[width=8cm]{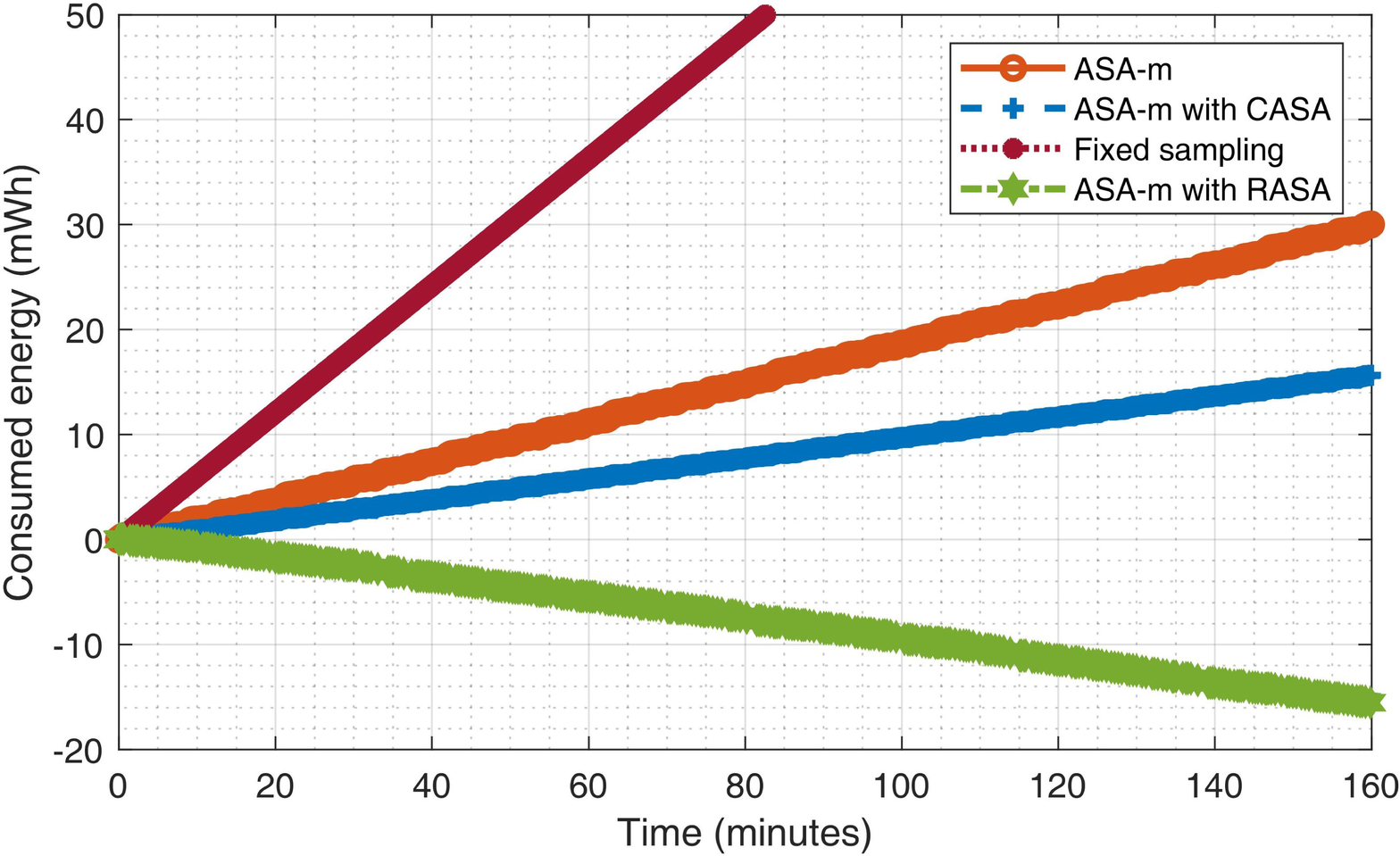}& \includegraphics[width=8cm]{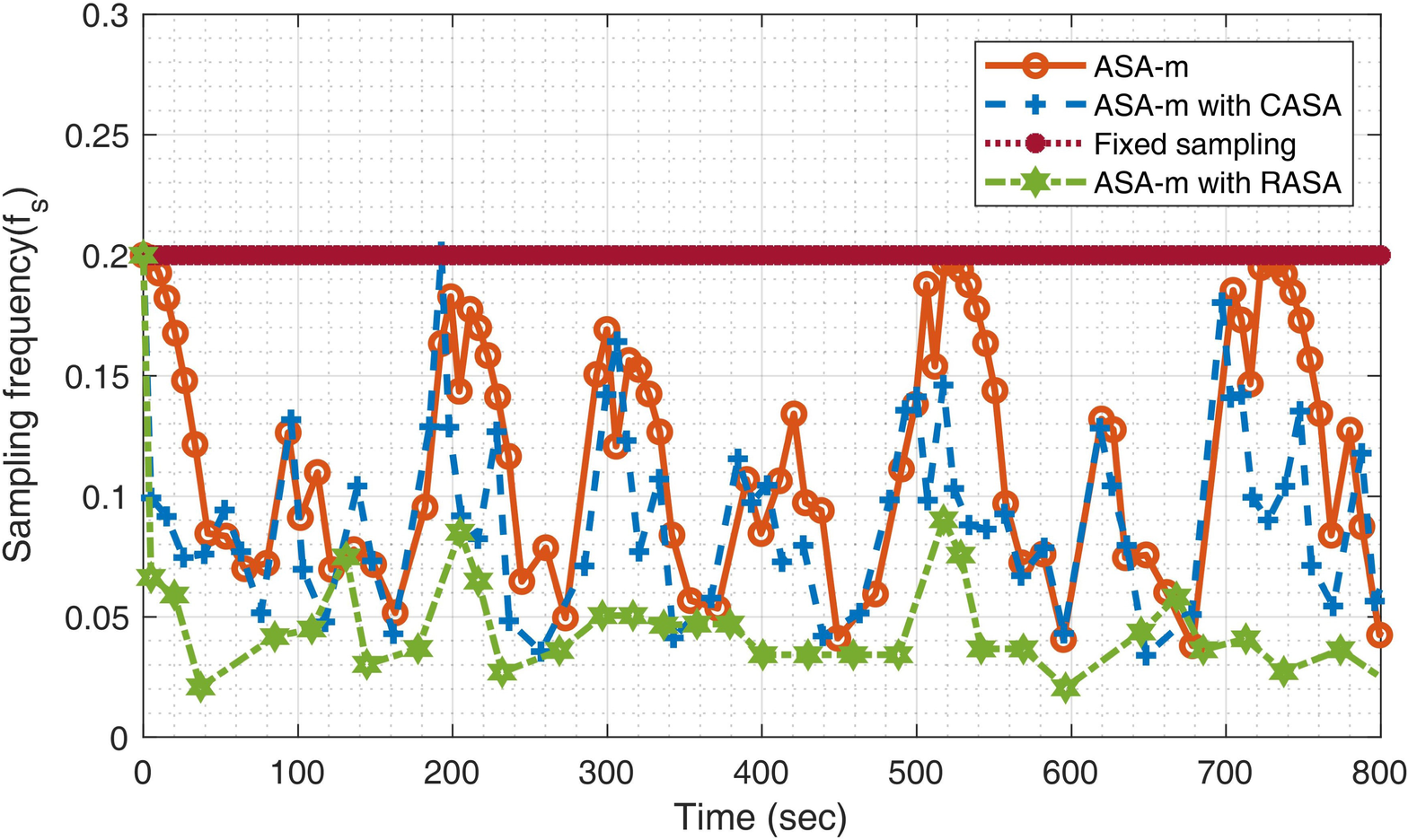}\\
(a) & (b)\\
\includegraphics[width=8cm]{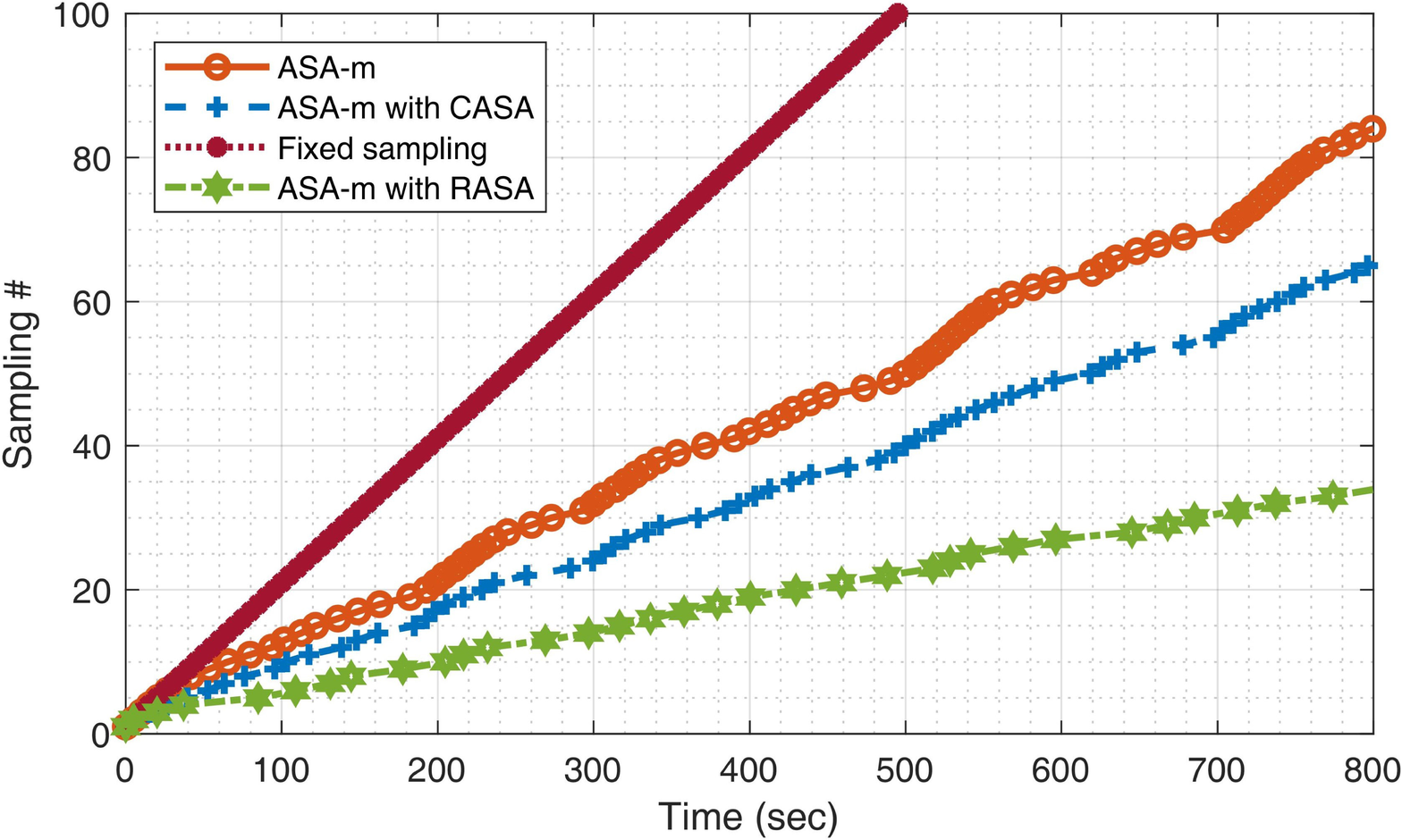}& \includegraphics[width=7.8cm]{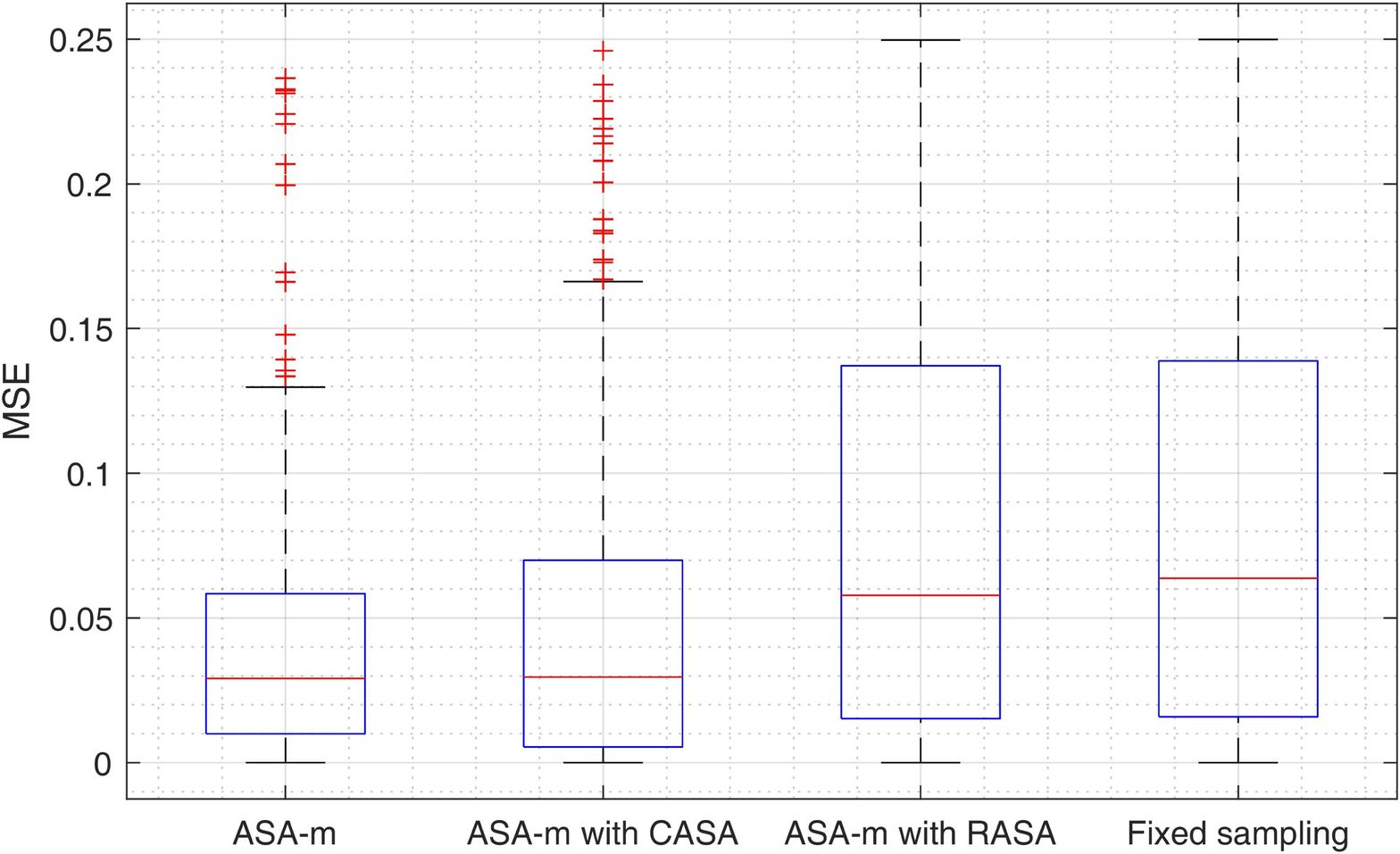}\\
(c) & (d)\\
\end{tabular}
\caption{The performance comparison of ASAs based on a dynamic transition data set.\label{ASA_dyna}}
\end{figure*}

\section{Simulation results}
\begin{figure*}[!t]
\centering 
\begin{tabular}{cc}
\includegraphics[width=8cm]{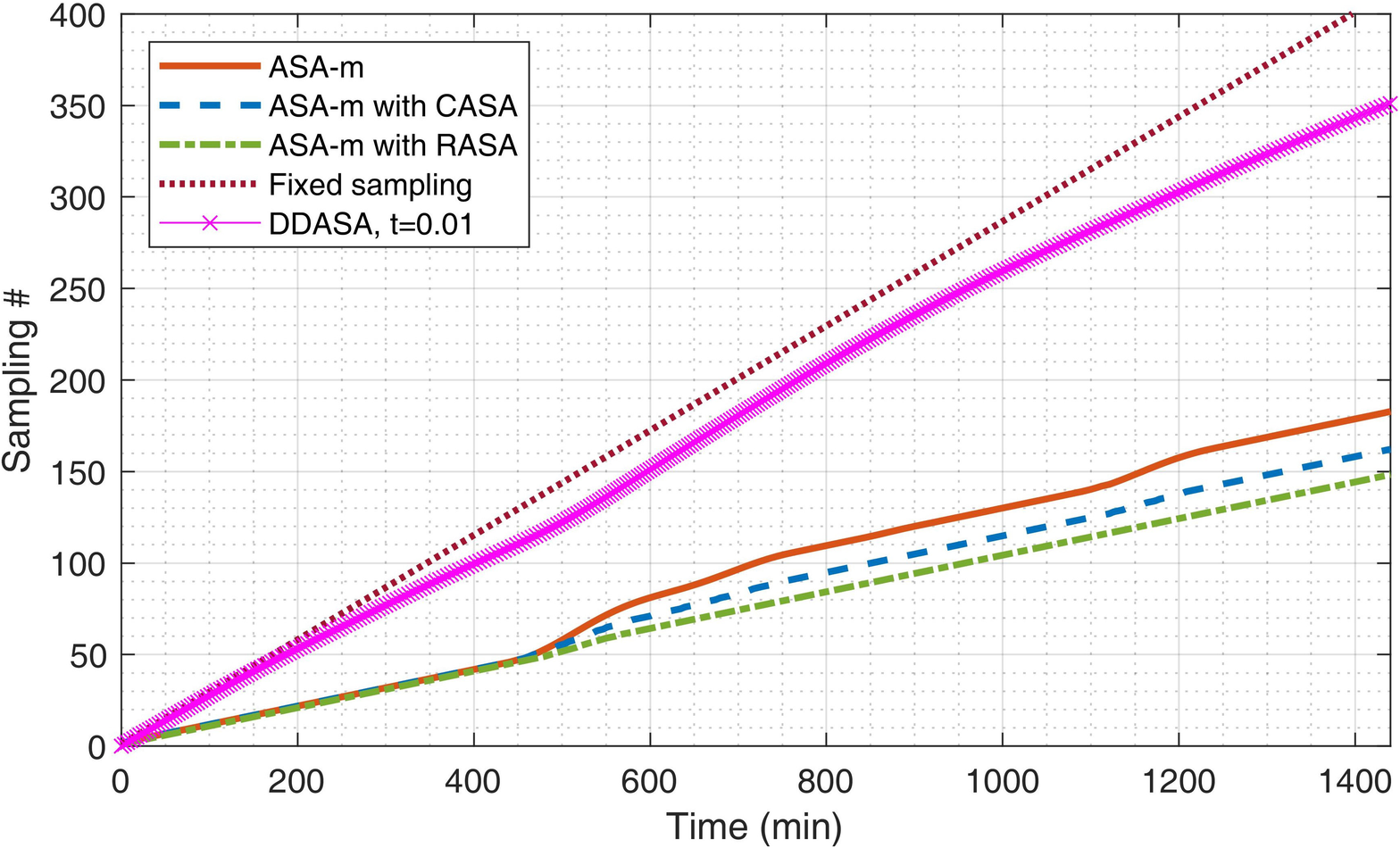}& \includegraphics[width=8cm]{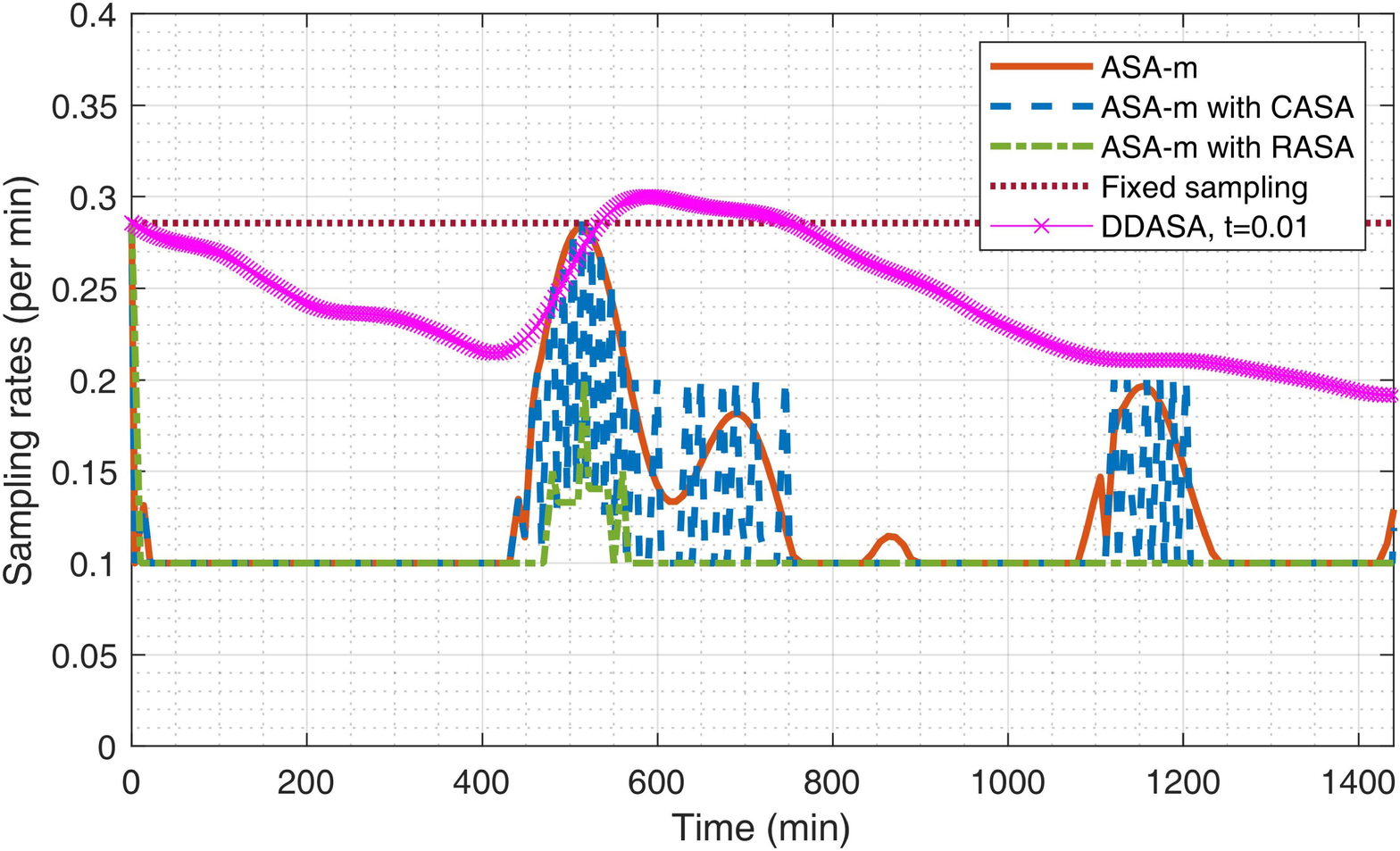}\\
(a) & (b)\\
\includegraphics[width=8cm]{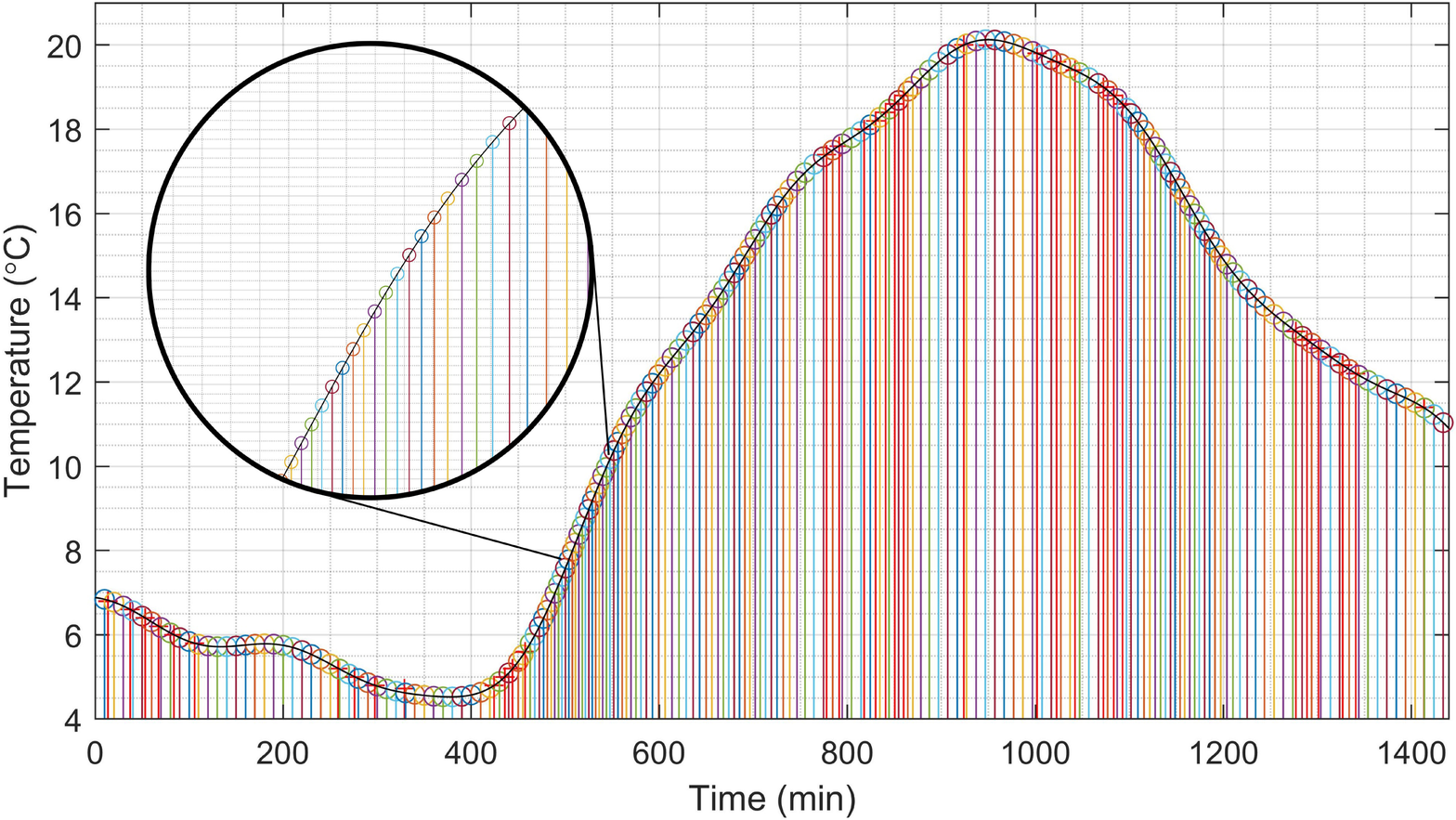}& \includegraphics[width=8cm]{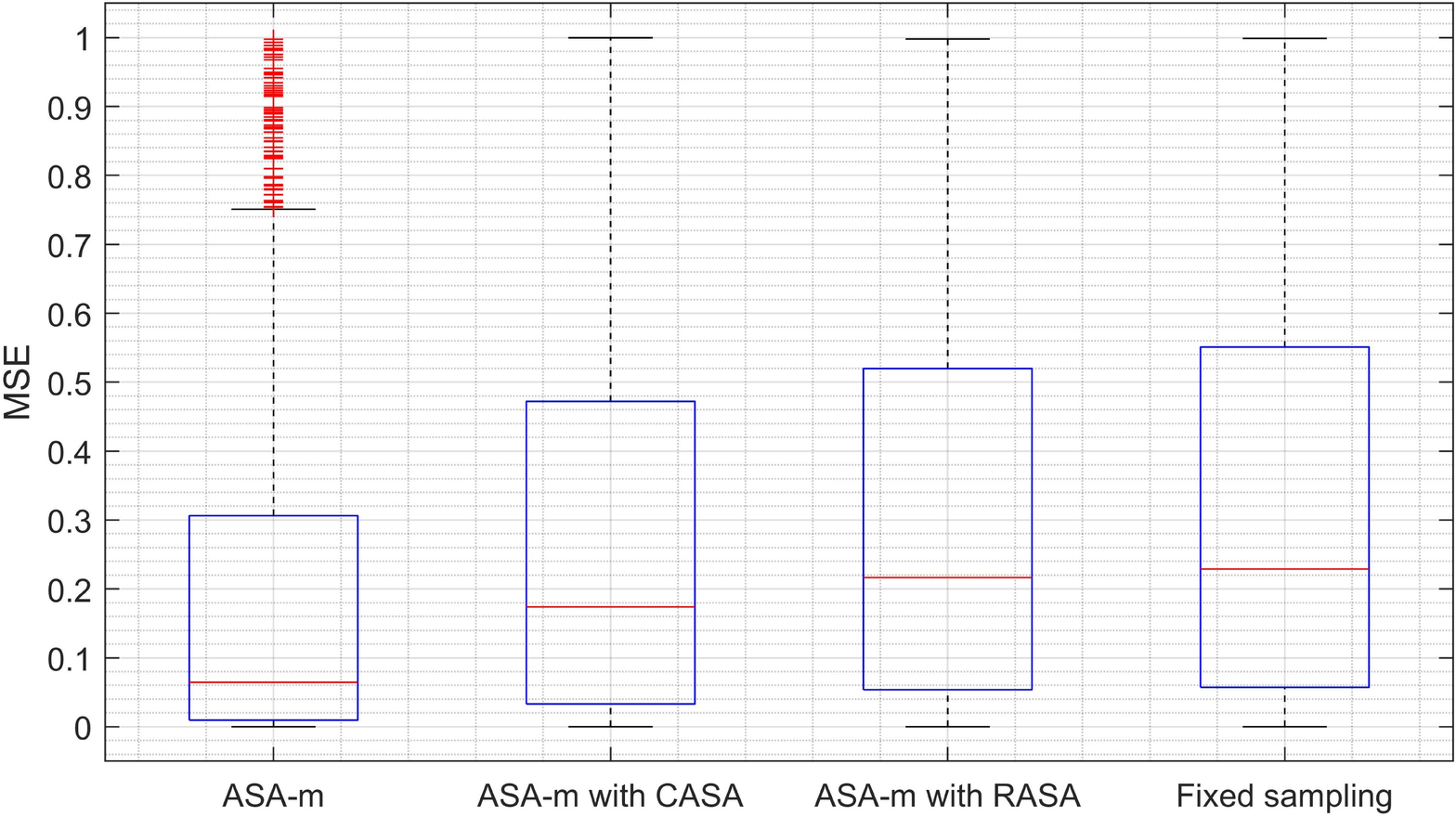}\\
(c) & (d)\\
\end{tabular}
\caption{The performance comparison of ASAs based on a KMA open access data during a day, 2020.04.13, in Seoul (Republic of Korea).\label{ASA_open}}
\end{figure*}

Figure~\ref{ASA}-\ref{ASA_dyna} show the results tested by using a dynamic transition data set and Figure~\ref{ASA_open} is the results tested based on a KMA open access data. Figure~\ref{ASA} compares the energy-saving performance of the proposed algorithms. This figure shows that RASA recharges the energy state of sensor nodes. It verifies the relation between a change of energy state and the count of sampled data by comparing Figure~\ref{ASA_dyna}(c) and~\ref{ASA}.

From Figure~\ref{ASA_dyna}(a), we observe the change of consumed energy states when the sensor node adapts the different algorithms. Comparing consumed energy states, we demonstrate the energy-saving performance of proposed algorithms. When the time is 82.5 minutes, fixed sampling consumed about $49.92mWh$. During the same time, ASA-m and CASA consumed about $17.27mWh$ and $8.32mWh$. ASA-m reduces the consumption energy up to $65.4\%$ compared with fixed sampling. CASA reduces the consumption energy up to $83.3\%$ and $54.5\%$ than fixed sampling and ASA-m. 

Figure~\ref{ASA_dyna}(b) compares the sampling frequencies by the proposed algorithms. It demonstrates that the sampling rate of the proposed algorithms does not consecutively reduce. It has a benefit when controlling sensor nodes. The sampling rate repeats actively increasing and decreasing according to the velocity of environmental change.

Figure~\ref{ASA_dyna}(c) shows the count of sampled data by proposed algorithms in ASMP. ASA-m saves up to $50\%$ of the sampling number than fixed sampling. CASA reduces up to $60\%$ of the sampling number than fixed sampling. Furthermore, RASA also saves up to $80\%$ the sampling number in the same conditions. In monitoring systems, the sampling rate is the most important factor related to energy consumption. This is because a sampling operation accompanies transmission and other processes. We demonstrate that RASA is the most energy-efficient sampling algorithm in Figure~\ref{ASA_dyna}(c).

Figure~\ref{ASA_dyna}(d) shows the error performance between a target ($B_n$) and sampled data ($D_n$). It demonstrates that ASA-m is the most accurate sampling algorithm. The fixed sampling has more errors than ASA-m because fixed sampling randomly gathers data without an estimation process. This causes a large gap between the target and the sampled data. MSE (Mean Squared Error) of ASA-m is $0.029$. ASA-m with CASA and RASA is $0.0296$ and $0.058$, respectively. All the proposed algorithms have greater performances than the fixed sampling, which had a value of $0.064$. This result reveals sampling accuracy. It means not missing or raising false alarms.

Figure~\ref{ASA_open} shows the simulation results of proposed and previous algorithms when they test based on KMA open access data set. That is a temperature data set during a day, 2020.04.13, in Seoul (Republic of Korea). That shows the advantages of ASMP in a real environment. These figures show the results of a fair comparison with a previous algorithm. Figure~\ref{ASA_open}(a)-(b) show the sampling advantage of our proposed algorithms. Proposed ASAs dynamically adjust the sampling rate among the energy-saving and transition-tracing. Through this ability, ASMP maximizes the energy-saving performance due to the fast restoration of the sampling rate. The error performance of DDASA is the same as that of fixed sampling because they collect a set of random data, no set of target data. 

\section{Conclusion}
As a result, our proposed protocol demonstrated the performance to satisfy all requirements for the most efficient disaster monitoring system, such as QoS, self-sustainability, and low calculation load. To arrange and test a disaster-monitoring system in real environments, self-sustainability is the most important factor because dead sensor nodes cause soil contamination. We propose a novel network protocol (ASMP) to ensure self-sustainability and QoS. 

ASMP comprises ASA-m, CASA, and RASA. ASA-m is an optimized sampling algorithm based on a novel adaptive filter, named MVP. ASA-m is advantageous to minimize unnecessary sampling while tracing environment changes autonomously. Using only ASA-m algorithm, the monitoring system highly improves its energy and system performance from the viewpoint of sustainability and QoS. CASA is an additional energy-saving technique that uses predicted data when a monitoring environment shows linear behavior. Using this property, CASA reduces energy consumption. RASA recharges the battery of sensor nodes when they face the danger of energy depletion. Energy-deficient sensor nodes have a chance to extend their lifetime using RASA.

Finally, we proved ASMP is the most effective sensor network protocol for EH-WSNs. ASMP immensely contributes to the build-up of reliable and self-sustaining EH-WSNs. Using two type data sets (i.e. Dynamic and Open Access), we conducted the reliable performance evaluations and clearly showed the advantages of our proposed works.

% or
%\appendix % for no appendix heading


%\appendices

%\section{}


\ifCLASSOPTIONcaptionsoff
 \newpage
\fi



\bibliographystyle{IEEEtran}
\bibliography{reference}
%\begin{thebibliography}{1}


%\end{thebibliography}


\begin{IEEEbiography}[{\includegraphics[width=1in,height=1.25in,clip,keepaspectratio]{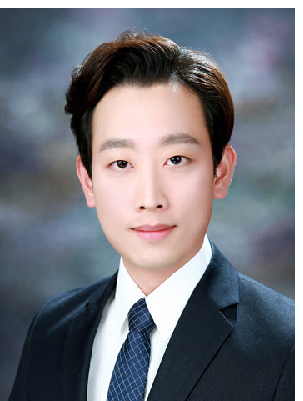}}]{Changmin Lee}
received a B.S. degree in Electronic Engineering from Chungnam National University, Daejeon, Republic of Korea in 2015, and a M.S. degree in Electrical and Electronic engineering from Yonsei university, Seoul, Republic of Korea in 2017. He participated in the work on open platform-based high-rise and complex facility multi-disaster countermeasure integrated CPS construction as a researcher of Multi Disaster Countermeasures Organization (MDCO), Korea Railroad Research Institute, Republic of Korea in 2019. He is currently undergoing a Ph.D. course in Yonsei university, Seoul, Republic of Korea.
\end{IEEEbiography}

\begin{IEEEbiography}[{\includegraphics[width=1in,height=1.25in,clip,keepaspectratio]{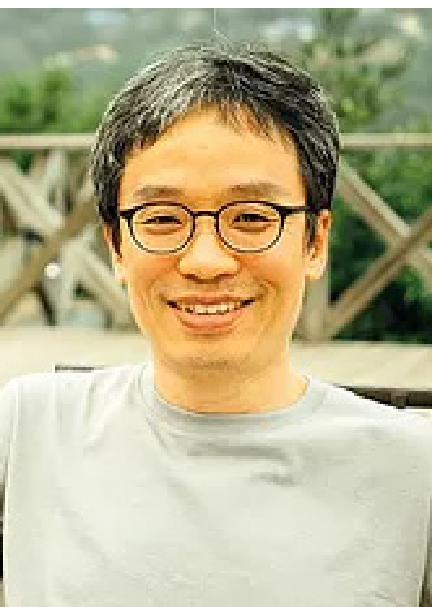}}]{Seong-Lyun Kim}
received a B.S. degree in Economics from Seoul National University, as well as M.S. and Ph.D. degrees in operations research with application to wireless networks from the Korea Advanced Institute of Science and Technology. He was an Assistant Professor of radio communication systems with the Department of Signals, Sensors, and Systems, Royal Institute of Technology (KTH), Stockholm, Sweden. He was a Visiting Professor with the Control Engineering Group, Aalto, Finland; the KTH Center for Wireless Systems; and
the Graduate School of Informatics, Kyoto University, Japan. He is currently a Professor of Wireless Networks with the School of Electrical and Electronic Engineering, Yonsei University, Seoul, Republic of Korea, and the Head of the Robotic and Mobile Networks Laboratory (RAMO) and the Center for Flexible Radio (CFR+). He is co-directing the H2020 EUK PriMO5G Project and leading Smart Factory TF of the 5G Forum, Republic of Korea. He has published numerous papers, including coauthoring the book, Radio Resource Management for Wireless Networks (with Prof. J. Zander). His research interests include radio resource management, information theory in wireless networks, collective intelligence, and robotic networks. He served as a Technical Committee Member, the Chair for various conferences, and as an Editorial Board Member for the IEEE TRANSACTIONS ON VEHICULAR TECHNOLOGY, the IEEE COMMUNICATIONS LETTERS, Elsevier Control Engineering Practice, Elsevier ICT Express, and the Journal of Communications and Network. He served as the Leading Guest Editor for the IEEE WIRELESS COMMUNICATIONS and the IEEE Network for wireless communications in networked robotics, and the IEEE JOURNAL ON SELECTED AREAS IN COMMUNICATIONS. He also consulted for various companies in the area of wireless systems in Republic of Korea and abroad.
\end{IEEEbiography}

\end{document}